\documentclass[preprint,secnumarabic,aps,prb]{revtex4}

\usepackage[tbtags]{amsmath}
\usepackage{amssymb,bm}
\usepackage{graphics}
\usepackage{color}

\newcommand{\mean}[1]{\left\langle#1\right\rangle}

\newcommand{\dd}{\mathrm{d}}
\newcommand{\br}{\bm{r}}
\renewcommand{\bf}{\bm{f}}
\newcommand{\bp}{\bm{p}}

\newcommand{\be}{\bm{e}}
\newcommand{\bR}{\bm{R}}
\newcommand{\bd}{\bm{d}}
\newcommand{\bK}{\bm{K}}
\newcommand{\bGamma}{\bm{\Gamma}}
\newcommand{\bpsi}{\bm{\psi}}

\newcommand{\gammal}{\gamma}
\newcommand{\gammat}{\overline{\gamma}}

\newcommand{\Wien}[2]{\int_{#1}^{#2}\!\bpsi(u)\,\dd u}
\newcommand{\Wieni}[3]{\int_{#1}^{#2}\!\bpsi_{#3}(u)\,\dd u}
\newcommand{\dpdWien}[3]{\Big[\int_{#1}^{#2}\!\psi_{#3}(u)\dd u\,\Big]}
\newcommand{\Liou}{\hat{\mathcal L}}
\newcommand{\Tav}{T_{av}}

\begin{document}

\title{Trotter derivation of algorithms for Brownian and dissipative particle
  dynamics} 
\author{Fabrice Thalmann$^{(1),}$\footnote{Author to whom correspondence 
should be addressed. Electronic mail thalmann@ics.u-strasbg.fr}
 and Jean Farago}
\affiliation{Institut Charles Sadron, CNRS UPR 
22, 67083 Strasbourg Cedex, France, and 
$^{(1)}$ Universit\'e Louis Pasteur, 4 rue
  Blaise Pascal, F-67070 Strasbourg Cedex France.}
\date{\today}

\begin{abstract}
  This paper focuses on the temporal discretization of the Langevin dynamics,
  and on different resulting numerical integration schemes. Using a method
  based on the exponentiation of time dependent operators, we carefully derive
  a numerical scheme for the Langevin dynamics, that we found equivalent to
  the proposal of Ermak and Buckholtz, \textit{J.Comput.Phys 35, p169 (1980)},
  and not simply to the stochastic version of the velocity-Verlet algorithm.
  However, we checked on numerical simulations that both algorithms give
  similar results, and share the same ``weak order two'' accuracy. We then
  apply the same strategy to derive and test two numerical schemes for the
  dissipative particle dynamics (DPD). The first one of them was found to
  compare well, in terms of speed and accuracy, with the best currently
  available algorithms.
\end{abstract} 
\maketitle

\section{Introduction}
\label{sec:Introduction}

Langevin dynamics is widely used in large-scale simulations of colloids,
polymers, and any system requiring a thermostat, either as a physical
ingredient of the system or as an expedient in very long simulations to
prevent the divergence of the temperature. This ubiquity of stochastic
dynamics in the realm of numerical simulations has motivated a lot of work to
design simple, efficient and accurate numerical schemes mimicking the Langevin
dynamics. For instance, in a recent paper~\cite{RicCic2003}, Ricci and Ciccotti
devised a discrete integration scheme for the Langevin dynamics of a set of
classical particles. Curiously enough, the resulting algorithm differs in a
way (apparently) difficult to reconcile with another well-known Langevin
algorithm, the so-called Ermak algorithm~\cite{Ermak,AllenTildesley}. In
particular, the Ermak algorithm needs two random numbers per degree of freedom
for each time step, whereas the Ricci-Ciccotti algorithm, which can also be
called stochastic Verlet algorithm, requires just one for an equal claimed
accuracy: as the computational effort is obviously different in each case,
either there is redundancy in one case, or an incorrect evaluation of the
precision of the algorithm in the other case.

One purpose of this paper is to explain where these differences stem from. To
this end, we revisit the techniques based on Trotter formulas in the context
of stochastic dynamics, and show that when commutators are taken into account
correctly up to the desired order, the Trotter formula yields the Ermak
algorithm.

However, we show also that the ``simplified'' algorithms using just one random
number per degree of freedom seem actually as good as the correct Ermak
algorithm, due to a statistical cancellation of the missing term, what the
numerical simulations amply confirm.

The second purpose of this paper follows naturally from these developments:
the technique based on a Trotter expansion can be easily adapted to the
description of the DPD dynamics, a stochastic dynamics \textit{\`a la}
Langevin which moreover preserves the momentum, and hence is extremely
appealing for applications requiring the stochasticity either as a tool to
stabilize the dynamics or as an effective ``heat bath'' in simulations of
non-equilibrium stationary states~\cite{dunweg}. As this dynamics was quite
recently introduced, not many algorithms have been proposed for the
DPD~\cite{shardlow,lesfinlandais}, some of them having been proved flawed (see
for instance the discussion in Ref.~\cite{2006_Pagonabarraga_Frenkel}). We
devise here a new DPD algorithm quite simple to implement and which has an
efficiency comparable to the Shardlow algorithm~\cite{shardlow}.

The paper is organized as follows: in the next section, we point out the
discrepancies which exist between the stochastic Verlet and the Ermak versions
of Langevin algorithms. Then, in Section~\ref{sec:MathematicalPreamble}, we
explain how to conduct a Trotter expansion in our Langevin context. Concluding
this part, we show why two random numbers per step and per degree of freedom
are, in principle, required in a simulation. The Langevin dynamics algorithms
resulting from our Trotter expansion are given in
Section~\ref{sec:LangevinDynamics}.  In a second time, we apply our
developments to the generation of DPD algorithms
(Section~\ref{sec:DissipativeParticleDynamics}) and discuss their efficiency
in Section~\ref{sec:NumericalDiscussion}. We show in particular that the
``bare'' efficiency of a DPD algorithm must be carefully disconnected from its
ability to handle the intrinsic singularity of the dissipation term when two
particles cross each other (a situation likely to happen only in a perfect gas
or with particles with soft-core potentials). In
Section~\ref{sec:NumericalDiscussion}, we also show numerics which endorse as
safe the use of stochastic velocity-Verlet algorithms, such as the one
proposed by Ricci and Ciccotti, in a Langevin simulation.

\section{The two Langevin dynamics discretization schemes}
\label{sec:twoSchemes}

The Langevin dynamics of a single particle with phase space coordinates
(position and momentum) $\br(t)$ and $\bp(t)$ reads:
\begin{equation}
\dot{\br}=\frac{\bp}{m}\,;\, 
\dot{\bp}=\bf(\br(t))-\gamma\bp(t)+\sigma\bm{\psi}(t), 
  \label{eq:basicLangevin}
\end{equation}
where $\bf$ is the external force acting on the particle (neighboring
particles and/or fields), $\gamma$ a friction coefficient (homogeneous to a
frequency), $T$ the temperature of the thermostat, \hbox{$\sigma
  =\sqrt{2m\gamma k_BT}$}, and $\bm{\psi}(t)$ a white noise with unit variance
(equivalently written as the derivative of a Wiener process)~: $\langle
\psi_\alpha(t)\rangle = 0$, $\langle
\psi_\alpha(t)\psi_\beta(t')\rangle=\delta_{\alpha,\beta}\delta(t-t')$
($\alpha$ and $\beta$ refer to the space indices).

In one version of Ermak scheme~\cite{remAllenTildesley}, velocities and
positions are updated according to the rule:  
\begin{eqnarray}
\br(t+\Delta t)  &=& \br(t) +\frac{1-e^{-\gamma\Delta t}}{\gamma}
\frac{\bp(t)}{m} +\frac{\Delta t^2}{2m}\bf(\br(t)) 
+\Delta \br^E; 
\label{eq:basicErmakVerlet:r}\\ 
\bp(t+\Delta t)  &=& e^{-\gamma\Delta t} \bp(t) 
  +\frac{1-e^{-\gamma\Delta t}}{2\gamma} 
\bigg[\bf(\br(t))+\bf(\br(t+\Delta t))\bigg]
+\Delta\bp^E,\label{eq:basicErmakVerlet:p}
\end{eqnarray}
where $\bp(t)$ and $\br(t)$ are the particle's impulsion and position at
time~$t$. In the absence of stochastic terms $\Delta \bp^E$ and $\Delta
\br^E$, and for vanishing friction coefficient $\gamma$, this algorithm
reduces to the well-known velocity-Verlet algorithm.

By integrating directly~(\ref{eq:basicLangevin}) between $0$ and
$\Delta t$, and neglecting terms of order higher than $\Delta t^2$, we obtain
the correlations between the two Gaussian increments $\Delta \br^E$ and
$\Delta \bp^E$~:
\begin{eqnarray}
\mean{\Delta \br^{E}} &=& \mathbf{0} \nonumber\\
\mean{\Delta \bp^{E}} &=& \mathbf{0} \nonumber\\
\mean{\Delta \br^{E}_\alpha\Delta \br^{E}_\beta} &=&
\frac{kT}{m\gamma^2} \bigg[2\gamma\Delta t -3 +4e^{-\gamma\Delta
  t}-e^{-2\gamma\Delta t}\bigg]
\delta_{\alpha,\beta}
\label{eq:CorrelationErmakVerlet}\\
\mean{\Delta \br^{E}_\alpha\Delta \bp^{E}_\beta} &=&
\frac{kT}{\gamma}\bigg[1-e^{-\gamma\Delta t}\bigg]^2 
\delta_{\alpha,\beta} \nonumber\\
\mean{\Delta \bp^{E}_\alpha\Delta \bp^{E}_\beta} &=& 
mkT \bigg[1-e^{2\gamma \Delta t}\bigg]\delta_{\alpha,\beta}.\nonumber
\end{eqnarray}
It is worth noting that $\Delta r^E$ is of order $\Delta t^{3/2}$ whereas
$\Delta p^E$ is of order $\Delta t^{1/2}$.

As regards the numerical scheme obtained by Ricci and
Ciccotti~\cite{RicCic2003},
\begin{eqnarray}
\br(t+\Delta t) &=& \br(t)+ \frac{\Delta t}{m} e^{-\gamma\Delta t/2}\bp(t) +
\frac{\Delta t^2}{2m}\bf(\br(t)) + \Delta\br^R;
\label{eq:RicciCiccotti:r}\\
\bp(t+\Delta t) &=& e^{-\gamma\Delta t}\bp(t) +\frac{\Delta t}{2}
\bf(\br(t+\Delta t)) e^{-\gamma\Delta t/4}+ \frac{\Delta
  t}{2}\bf(\br(t))e^{-3\gamma\Delta t/4}
+\Delta\bp^R,\label{eq:RicciCiccotti:p}
\end{eqnarray}
it is equivalent to~(\ref{eq:basicErmakVerlet:r})
and~(\ref{eq:basicErmakVerlet:p}) up to order $\Delta t^2$, except for the
random increments $\Delta \br^R$ and $\Delta \bp^R$ which show slightly
different correlations~:
\begin{eqnarray}
\mean{\Delta \bp^{R}} &=& \mathbf{0} \nonumber\\
\mean{\Delta \bp^{R}_\alpha\Delta \bp^{R}_\beta}
&=& 
\sigma^2 e^{-\gamma\Delta t}\Delta t\delta_{\alpha,\beta}
\label{eq:RicciCiccotti13}\\
\Delta\br^R &=& e^{\gamma\Delta t/4}\frac{\Delta t}{2m}\Delta\bp^R
\simeq\frac{\Delta t}{2m}\Delta\bp^R.\nonumber   
\end{eqnarray}
%
%
In~(\ref{eq:RicciCiccotti13}), the increment $\Delta \br^R$ is proportional to
$\Delta \bp^R$, and the algorithm requires only the generation of a single
random number per step. \\ 

Because both algorithms aim at reaching the same accuracy, or at least the
same order in a power expansion in $\Delta t^n$, it is important to understand
the difference between the statistics of the random increments. To
this end, a natural tool is the Trotter splitting technique, as it
was invoked in Ref.~\cite{RicCic2003}.


\section{Mathematical preamble}
\label{sec:MathematicalPreamble}

\subsection{Algebraic formulation of the Langevin dynamics}

The road to finding efficient Brownian numerical algorithms, passes by
reformulating this dynamical problem in the language of operators.  Once
expressed in this language, the mathematics provide a number of convenient
rules, such as the Trotter, or the Baker-Campbell-Hausdorff (BCH) formulas, to
handle these non-commuting operators.

This process can be illustrated on a single particle autonomous equation of
motion, with a mass $m$ set arbitrarily to~1, without loss of generality:
\begin{equation}
\dot \br=\bp;\; \dot \bp = \bf(\br).\label{eq:deterministic3D}
\end{equation}
The resulting dynamics can be subject to a double point of view. The most
usual one consists in following the phase-space trajectory $(\br(t),\bp(t))$
of the particle as a function of the time $t$ and the initial condition
$(\br_0,\bp_0)$. Drawing an analogy with fluid mechanics, this corresponds to
a kind of Lagrangian point of view, where the dynamics consists in a mapping
of the phase-space on itself:
\begin{equation} 
(\br(t),\bp(t)) = \Phi(\br(0),\bp(0);t),
\end{equation}
commonly known as the ``Hamiltonian flow''. The mapping $\Phi$ is a time
evolution operator acting on the phase-space coordinates.  This mapping is
sometimes loosely symbolized as $(\br(t),\bp(t)) =e^{Lt}(\br(0),\bp(0))$ but
this notation can be misleading as $L$ acts on the phase space coordinates,
and must not be confused with the Liouvillian operator introduced below.

The other, dual, point of view consists in focusing on the observables
$\rho(\br,\bp;t)$ of the phase space, for a fixed reference point $(\br,\bp)$.
This approach is essentially similar to the Eulerian description of a flow.
For instance, if we consider a probability density function (pdf)
$\rho(\br,\bp;t)$ describing the motion of a particle subject to the
dynamics~(\ref{eq:deterministic3D}), then the pdf taken at two different times
$t=0$ and $t$ are connected by the relation~:
\begin{equation}
 \rho(\br,\bp;t) = \rho(\br',\bp';0), \label{eq:observableDeterministic3D}
\end{equation}
where $(\br',\bp')$ is the reciprocal point of $(\br,\bp)$ by the mapping
$\Phi$: 
\begin{align}
 &(\br,\bp)= \Phi(\br',\bp';t) \text{ or } (\br',\bp')= \Phi^{-1}(\br,\bp;t).
\end{align}
Moreover, for a stationary dynamics, the inversion of $\Phi$ is just a time
reversal, leading to $(\br',\bp')= \Phi(\br,\bp;-t)$. Specifying
(\ref{eq:observableDeterministic3D}) to an infinitesimal time step $\dd t$, we
obtain the Liouvillian dynamics of the observable $\rho(\br,\bp;t)$
\begin{eqnarray}
\frac{\partial \rho}{\partial t} &=& \left(-\bp\cdot\frac{\partial}{\partial
    \br}-\bf(\br)\cdot\frac{\partial}{\partial \bp}\right) \rho;
\label{eq:LiouvillianDeterministic}\\
&=& -\Liou \rho,\notag
\end{eqnarray}
with $\Liou = \bp\cdot\partial/\partial \br+\bf\cdot\partial/ \partial \bp$,
the Liouvillian operator. The partial differential
equation~(\ref{eq:LiouvillianDeterministic}) can be integrated with respect to
time, and the pdf $\rho(\br,\bp;t)$ expressed in terms of its initial
condition:
\begin{equation}
\rho(\br,\bp;t) = \exp(-t\Liou)\rho(\br,\bp;0).
\label{eq:LiouvillianExponential}
\end{equation}
The time evolution operator $\exp(-t\Liou)$ is a well defined mathematical
object, belonging to an operator algebra acting on the Hilbert space of
differentiable functions of phase space, such as $\rho(\br,\bp;t)$. Basically,
$\exp(-t\Liou)$ carries the same amount of information as the Hamiltonian
flow, and its knowledge amounts to the knowledge of the mapping $\Phi$.  An
intuitively appealing choice for $\rho$ is the narrow, bell shaped function
\hbox{$\rho_{\epsilon}=\exp(-[(\br-\br_0)^2+(\bp-\bp_0)^2]/2\epsilon^2)$} of
width $\epsilon$.  Plotting $\rho_{\epsilon}(\Phi(\br,\bp;-t))$ as a function
of $r$ and $p$ for various successive times will show the time evolution of a
``packet'' of independent particles with initial position close to
$(r_0,p_0)$, each one following its own Hamiltonian trajectory. In higher
dimensional phase-spaces with deterministic chaos, the packet spreads
progressively and eventually occupies all the permitted subspace of positions
compatible with conserved quantities (ergodicity). The $\epsilon\rightarrow
0$ limit of this corresponds to an initial condition
$\rho(\br,\bp,0)=\delta(\br-\br_0)\delta(\bp-\bp_0)$; it is easy to verify
that $\rho(\br,\bp,t)=\delta(\br-\br(t))\delta(\bp-\bp(t))$, where $\br(t)$
and $\bp(t)$ are the solutions of the equations of motion with initial
condition $(\br_0,\bp_0)$; this shows the complete equivalence between the
``coordinate'' and the ``observable'' version of the dynamics.

In our derivation of discrete integration schemes for the Langevin dynamics,
we adopt this second point of view, consisting of operators acting on the
functions of phase space, and we stress upon the importance of keeping the
minus sign when defining the time evolution operator
in~(\ref{eq:LiouvillianExponential}). Provided this is correctly taken into
account, many convenient algebraic techniques apply, while getting the same
results with the point of view of flows acting on coordinates would simply be
awkward. Moreover, this convention resembles the familiar operator calculus
of quantum mechanics (the observable $\rho(\br,\bp;t)$ playing the role of
wave-function).  We leave to appendix~\ref{app:sign} a simple example showing
why working on observables gives better results than working with coordinates.

\subsection{The Velocity-Verlet algorithm}

For a stationary dynamics, the next step is to split the time interval $t$ into
$n$ small steps $\Delta t$ and write:
\begin{equation}
\exp(-t\Liou) = \exp(-\Delta t\Liou)^n,
\end{equation}
which simply reduces the dynamics to the iteration of a large number of
identical elementary steps, each one representing the dynamics on a time step
$\Delta t$. Finding a good integration scheme is equivalent to finding a good
estimate of $\exp(-\Delta t\Liou)=\exp(A+B)$, with
\hbox{$A=-\Delta t\,\bf(\br)\cdot\partial/\partial\bp$} and 
\hbox{$B=-\Delta t\,\bp\cdot\partial/\partial\br$}. 
Like in many cases, the complexity of the dynamics comes from the fact that
for two non-commuting operators $A$ and $B$, $\exp(A+B)$ is not
$\exp(A)\exp(B)$, and even though $\exp(A)$ and $\exp(B)$ are known, the
exponential of the sum $\exp(A+B)$ cannot be calculated. To overcome this
difficulty, the strategy is to reduce the norm of $A$ and $B$ until their
commutator $[A,B]=AB-BA$ becomes small enough to apply the
Baker-Campbell-Haussdorf (BCH) formula, stating that:
\begin{eqnarray}
\exp(A)\exp(B)&=&\exp\left(A+B+\frac{1}{2}[A,B]+\frac{1}{12}[A,[A,B]]
+\frac{1}{12}[B,[B,A]]+R\right),
\label{eq:BCHDirect3}
\end{eqnarray}
with a remaining $R$ consisting of a sum of commutators involving more than
four operators $A$ or $B$ each. In the present case, both $A$ and $B$ scale as
$\Delta t$, which ensures that for a time step small enough, $R$ is just a
negligible perturbation term. In practice, the BCH is reversed to take the
form
\begin{equation}
\exp(A+B)=\exp(A)\exp(B)\exp\left(\frac{1}{2}[B,A]-\frac{1}{3}[B,[B,A]]
+\frac{1}{6}[A,[A,B]]+R'\right),
\label{eq:BCHReverse3}
\end{equation}
and one checks that 
\begin{equation}
\exp(A+B)=\exp(A/2)\exp(B)\exp(A/2)\exp\left(-\frac{1}{12}[B,[B,A]]
+\frac{1}{24}[A,[A,B]]+R''\right),
\label{eq:BCHReverse3:2}
\end{equation}
with $R'$ and $R''$ of order $\Delta t^4$. Thus, the two following relations
hold, up to terms of order $\Delta t^2$.
\begin{eqnarray}
\exp(-\Delta t\Liou)&=& 
  \exp\left(-\frac{\Delta t}{2}\bp\cdot\frac{\partial}{\partial \br}\right) 
  \exp\left(-\Delta t\,\bf(\br)\cdot\frac{\partial}{\partial \bp}\right)
  \exp\left(-\frac{\Delta t}{2}\bp\cdot\frac{\partial}{\partial \br}\right);
\label{eq:splittingVV:1}\\
&=& \exp\left(-\frac{\Delta t}{2}\bf(\br)
  \cdot\frac{\partial}{\partial \bp}\right) 
  \exp\left(-\Delta t\bp\cdot\frac{\partial}{\partial \br}\right)
  \exp\left(-\frac{\Delta t}{2}\bf(\br)\cdot\frac{\partial}{\partial
  \bp}\right).
\label{eq:splittingVV:2}
\end{eqnarray}
As recalled in appendix~\ref{app:vv}, both splitting are equivalent to
the Verlet algorithm.

\subsection{The case of a non stationary dynamics}
\label{subsec:nonStatDynamics}
When the dynamics is not stationary, the Liouvillian is time-dependent, and
the evolution operator can be expressed as a time-ordered, or chronological
exponential:

\begin{equation}
\begin{split}
\mathrm{T}\exp\left(-\int_{t_1}^{t_2}\dd s \Liou(s)\right)
&= 1 - \int_{t_1}^{t_2} \dd s\, \Liou(s) + \int_{t_1}^{t_2}\dd
s_1\int_{t_1}^{s_1}\dd s_2\,\Liou(s_1)\Liou(s_2) \ldots\\
&= \sum_{n=0}^{\infty} \frac{(-1)^n}{n!}\mathrm{T}
\left(\int_{t_1}^{t_2} \dd s \Liou(s)\right)^n,
\end{split}
\label{eq:T-expDefinition}
\end{equation}
The symbol~T ensures that all the time dependent operators are written
from right to left in chronological order. Chronological exponentials must be
used as soon as the differential operators $\Liou(s)$ and $\Liou(s')$ taken at
two different times, are found to be non-commuting. This situation indeed
arises when considering the Langevin dynamics of
equation~(\ref{eq:basicLangevin}):
\begin{equation}
\dot \br=\frac{\bp}{m};\; 
\dot \bp = \bf(\br) -\gamma \bp +\sigma \bpsi(t).
\label{eq:stochastic1D} 
\end{equation}
The equation for the momentum is modified by a friction coefficient $-\gamma
\bp$ and a random (or Langevin) force $\bpsi(t)$, obeying the usual fluctuation
dissipation relation. Straightforwardly, we associate to this stochastic
equation a ``Liouvillian'':
\begin{equation}
\Liou(t) = \frac{\bp}{m} \cdot\frac{\partial}{\partial \br} 
 +\left(\bf(\br)-\gamma \bp
 +\sigma\bpsi(t) \right)\cdot\frac{\partial}{\partial \bp}
\label{eq:LiouvilleStochastic},
\end{equation}
where the time dependence lies in the random force.
By doing so, we implicitly assume a Stratonovitch interpretation of the
stochastic equation, the only one for which the stochastic equivalent of 
eq.~(\ref{eq:LiouvillianDeterministic}) holds~\cite{RemarkIto}.

The handling of $\mathrm{T}\exp(-\int_t^{t+\Delta t}\Liou(s))$ shows important
differences with the autonomous, time-independent Hamiltonian situation.
First, one cannot simply replace $\mathrm{T}\exp(-\int_t^{t+\Delta
  t}\Liou(s))$ by an ordinary exponential without possibly introducing an
error of order $\Delta t^{3/2}$. Second, when splitting the Liouvillian into
$A=\Delta t\,\bp/m\cdot\partial/\partial \br$ and $B=\Liou-A$, the term $B$ is
of order $\Delta t^{1/2}$, due to the presence of the random force. This
entails that the triple commutator $[B,[A,B]]$ in the BCH formula can be of
order $\Delta t^2$. It is essential to either check that this commutator
vanishes (as for the usual Langevin dynamics), or to keep it in the
calculation (as for instance in the DPD situation).

The basic tool for handling the T-exponential is the Magnus
expansion~\cite{BurBur99}, which in the present case reads:
\begin{multline}
\displaystyle
\exp\bigg(\int_t^{t+\Delta t} \dd s\, \Liou(s)\bigg)\cdot
\mathrm{T}\exp\bigg(-\int_t^{t+\Delta t} \dd s\, \Liou(s)\bigg)=\\
\displaystyle
\exp\bigg(-\frac{1}{2}\int_{t}^{t+\Delta t}\!\dd s\int^{\displaystyle s}_{t} 
 \dd u\, [\Liou(u),\Liou(s)]\\
-\frac{1}{3} \int_{t}^{t+\Delta t}\!\dd s
 \int_t^{\displaystyle s} \dd u \int_t^{\displaystyle s} \dd v\,
 [\Liou(v),[\Liou(u),\Liou(s)]]\bigg),
\label{eq:TExpTripleCommutator}
\end{multline}
We provide in appendix~\ref{app:Magnus} a simple demonstration
of~(\ref{eq:TExpTripleCommutator}). It is clear that the right hand side
reduces to the operator identity for a time independent Liouvillian.
Using~(\ref{eq:TExpTripleCommutator}), one can replace the chronological
exponential by a simple exponential, which can itself be simplified using a
BCH formula with triple commutators~(\ref{eq:BCHReverse3}).

Finally, let us notice that rather than dealing directly with the stochastic
differential equation (s.d.e), it is also possible to go to the Fokker-Planck
(FP) formulation of the stochastic process.  In that case, the dynamics
expresses naturally in terms of Fokker-Planck operators, for which a similar
BCH decomposition strategy can be done~\cite{FabSerEspCov2006}, and a
numerical scheme can be obtained~\cite{SerFabEspCov2006}. In the FP
approach, there is no time-dependent term in the evolution operator and it is
not necessary to consider the time commutators appearing
in~(\ref{eq:TExpTripleCommutator}), but any connection with the
underlying stochastic noise is lost in the process.


\subsection{Application to Langevin dynamics}
\label{subsec:ApplicationLangevin}

We now focus in detail on~(\ref{eq:stochastic1D}).
The evolution operator can be rewritten as
\begin{multline}
\mathrm{T}\exp\left(-\int_t^{t+\Delta t} \dd s\, \Liou(s)\right)=\\
\exp\left(-\int_t^{t+\Delta t} \dd s\, \Liou(s)\right)
 \cdot \exp\left(\int_t^{t+\Delta t} \dd s\, \Liou(s)\right)\cdot
\mathrm{T}\exp\left(-\int_t^{t+\Delta t} \dd s\, \Liou(s)\right).
\end{multline}
with the product of the two last terms given by the
identity~(\ref{eq:TExpTripleCommutator}). Fortunately in this case, the triple
commutators give rise only to terms of order~$\Delta t^{5/2}$, which can be
neglected as we aim only at a $\Delta t^2$ accuracy. Thus, 
\begin{multline}
\mathrm{T}\exp\left(-\int_t^{t+\Delta t} \dd s\, \Liou(s)\right)
= \exp\left(-\int_t^{t+\Delta t} \dd s\, \Liou(s)\right)\\
\times\exp\left(-\frac{1}{2}\int_t^{t+\Delta  t}\dd s\,\int_0^{\displaystyle
  s}\dd u\, [\Liou(u),\Liou(s)] \right).
\label{eq:decompositionEvolution}
\end{multline}
Integrating by part, we get
\begin{multline}
-\frac{1}{2}\int_t^{t+\Delta t}\!\dd s\,\int_t^{\displaystyle s}\dd u\, 
[\Liou(u),\Liou(s)] \\
= \sigma\left[\frac{\Delta t}{2}\Wien{t}{t+\Delta t}
  -\int_t^{t+\Delta t}  \dd  s\,\Wien{t}{s}\right]
  \left(\frac{1}{m}\frac{\partial}{\partial r}  
  -\gamma \frac{\partial}{\partial p}\right). 
\label{eq:contentDoubleCommutatorLangevin}
\end{multline}
This time-commutator accounts precisely for the difference between the
algorithm of Ricci and Ciccotti and the algorithm of Ermak.  To support this
claim, we simply compare up to order $\Delta t^2$, the exact time-evolution
operator $\mathrm{T}\exp(-\int_t^{t+\Delta t}\dd s\Liou(s))$, and the
corresponding simple exponential $\exp(-\int_t^{t+\Delta t}\dd s\Liou(s))$, in
the case of the s.d.e~(\ref{eq:basicLangevin}).
A direct calculation, in the exact case, at the order $\Delta t^2$, gives:
\begin{eqnarray}
\br(t+\Delta t) &=&  \br(t)+\bp(t) \frac{\Delta t}{m} + \Big(\bf(\br(t)) 
-\gamma\bp(t)\Big)\frac{\Delta t^2}{2m}+ \frac{\sigma}{m}\int_0^{\Delta t}\!\dd
s\,\Wien{t}{s};\nonumber\\ 
\bp(t+\Delta t) &=& \bp(t) + \Big(\bf(\br(t))-\gamma \bp(t)\Big) \Delta t +
\Big(\frac{\bp(t)}{m}\cdot\nabla_r\bf(r(t)) -\gamma \bf(\br(t))
+\gamma^2\bp(t)\Big) \frac{\Delta t^2}{2} 
\nonumber\\
& & +\sigma\left\lbrace\Wien{t}{t+\Delta t}-\gamma\int_t^{t+\Delta t}\!\dd
  s\,\Wien{t}{s}\right\rbrace,  
\label{eq:exactSolutionOrder2}
\end{eqnarray}
and is equivalent to applying~(\ref{eq:decompositionEvolution}), and to Ermak.
The simple exponential case $\exp(-\int_t^{t+\Delta t}\dd s\Liou(s))$ amounts
to replacing the Langevin force by a stepwise averaged quantity, and to
redefining the force accordingly:
\begin{equation}
\bf(\br) \mapsto \tilde{\bf}(r)= \bf(\br) +\frac{\sigma}{\Delta
  t}\int_t^{t+\Delta t} \bpsi(s)\dd s.  
\end{equation}
It gives:
\begin{eqnarray}
\br(t+\Delta t) &=&  \br(t)+\bp(t)\frac{\Delta t}{m} +
\Big(\bf(\br(t))-\gamma\bp(t)\Big)\frac{\Delta t^2}{2m}+ 
\frac{\sigma}{m}\frac{\Delta t}{2}\Wien{t}{t+\Delta t};\\ 
\bp(t+\Delta t) &=& \bp(t) + \Big(\bf(\br(t))-\gamma \bp(t)\Big)\Delta t +
 \Big(\frac{\bp(t)}{m}\cdot\nabla_r\bf(r(t)) -\gamma\bf(\br(t))
+\gamma^2 \bp(t)\Big) \frac{\Delta t^2}{2}\nonumber\\ 
 & & +\sigma\Wien{t}{t+\Delta t}
-\gamma\sigma\frac{\Delta t}{2}\Wien{t}{t+\Delta t}.
\label{eq:averagedSolutionOrder2}
\end{eqnarray}
Expression~(\ref{eq:averagedSolutionOrder2}) is equivalent to the stochastic
Verlet or Ricci and Ciccotti result.
Expressions~(\ref{eq:exactSolutionOrder2}) and
(\ref{eq:averagedSolutionOrder2}) differ by a term $\mathrm{O}(\Delta
t^{3/2})$, which turns out to be equal to the quantity appearing
in~(\ref{eq:contentDoubleCommutatorLangevin}):
\begin{eqnarray}
\left[\frac{\Delta t}{2}\Wien{t}{t+\Delta t}-\int_t^{t+\Delta t} 
\dd s\,\Wien{t}{s}\right] 
=\int_t^{t+\Delta t} \left(s-\frac{\Delta t}{2}\right)\bpsi(s)\dd s. 
\label{eq:uncorrelatedTerm}
\end{eqnarray}
We note that, in order to get the correct result by using the usual commutator
algebra, it is absolutely necessary to use the correct convention for the
Liouvillian, \textit{i.e.}  $\exp(-t\Liou)$ for the forward dynamics of the
observables. The above equality is obtained by integrating by part the Wiener
process. Thus, the Gaussian increments arising
in~(\ref{eq:exactSolutionOrder2}) can be split in two statistically
independent parts:
\begin{eqnarray}
\Delta r^E &=& \frac{\sigma}{m}\frac{\Delta t}{2}\int_t^{t+\Delta
 t}\bpsi(s)\dd s 
 +\frac{\sigma}{m}\int_t^{t+\Delta t}\left(\frac{\Delta t}{2}-s\,\right)
  \bpsi(s)\dd s;\\
\Delta p^E &=& \sigma\bigg(1-\frac{\gamma\Delta t}{2}\bigg)
\int_t^{t+\Delta t}\bpsi(s)\dd s  
-\gamma\sigma\int_t^{t+\Delta t} 
\left(\frac{\Delta t}{2}-s\,\right) \bpsi(s)\dd s,
\end{eqnarray}
each part having zero mean  and correlations:
\begin{eqnarray}
\mean{\left(\int_t^{t+\Delta t}\bpsi(s)\dd s\right)^2} &=& 
\Delta t;\\ 
\mean{\left(\int_t^{t+\Delta t} \left(\frac{\Delta t}{2}-s\,\right)\bpsi(s)\dd
    s\right)^2} &=& \frac{\Delta t^3}{12};\\ 
\mean{\int_t^{t+\Delta t}\bpsi(s)\dd s
\int_t^{t+\Delta t} \left(\frac{\Delta t}{2}-s\,\right)\bpsi(s)\dd s}
&=& 0. \label{eq:ErmakNoCorrelation}
\end{eqnarray}
The statistical independence~(\ref{eq:ErmakNoCorrelation}) is
crucial.  It ensures that the difference between
(\ref{eq:exactSolutionOrder2}) and (\ref{eq:averagedSolutionOrder2}), or
equivalently the correction term brought
by~(\ref{eq:contentDoubleCommutatorLangevin}) contributes only at the order
$\Delta t^3$ in variance. The current algorithm being a weak order-two
algorithm, this term is not relevant and negligible at this order. We conclude
from this calculation that, when taking properly into account all the terms
potentially of order $\Delta t^{3/2}$, the algorithm obtained is Ermak.
However, (\ref{eq:uncorrelatedTerm})~turns out to be uncorrelated with the
dominant contribution of the random force $\int\bpsi(s)\dd s$, which in turn
guarantees that the correction is no longer relevant for a weak order-two
algorithm.  Nonetheless, this statement is not obvious, and the
conclusion can only be drawn once the corrective terms have been evaluated.

\begin{figure}[ht]
\begin{center}
 \rotatebox{-90}{\resizebox{0.8\textheight}{!}{\includegraphics*{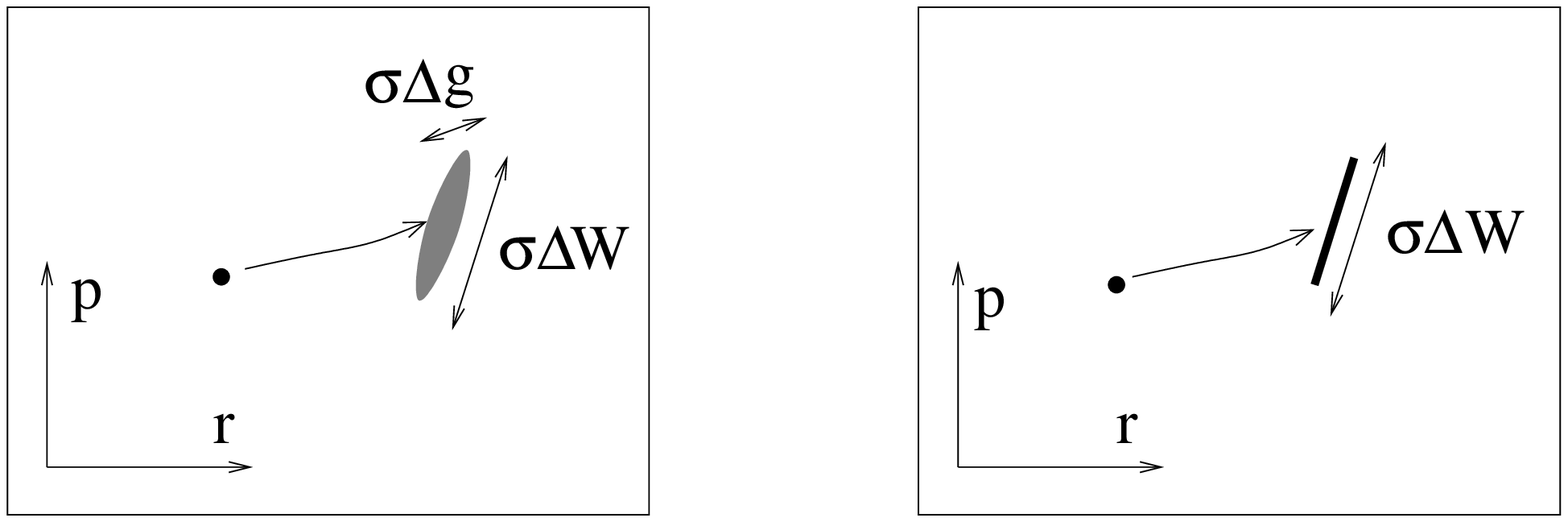}}}
\caption{} 
\label{fig:riceGrain}
\end{center}
\end{figure}

We conclude by giving a pictorial representation of the difference of these
two algorithms. From a given noise realization, the stochastic Verlet
algorithm gets rid of half of the random numbers that the Ermak algorithm uses,
and in a sense, throws away an equal amount of information. On the other hand,
the generation of the $\Delta g=\int(s-\Delta t/2)\bpsi(s)\dd s$ may seem
futile, as this is not real information which is ``created'', but just pure
randomness.  We believe that the issue of this paradoxical situation is that
the generation of $\Delta g$ indeed makes sense, even though it is not proper
information.  To support this claim, we come back to the Fokker-Planck (FP)
representation of the stochastic process. In the FP representation, one aims
at calculating the propagator associated to the elementary time interval:
\begin{equation}
\mathcal{P}(r,p;r_0,p_0;\Delta t) = \exp(\Delta t \Liou_{FP})\,
\delta(p-p_0)\delta(r-r_0).
\end{equation}
This function is a bell shaped curve, which is close to a Gaussian
distribution. In the two dimensional space $(r,p)$, this function is centered
around the position that would occupy the particle in the absence of noise,
and has a ``rice grain'' shape, with a large axis oriented along the $p$ axis,
parallel to the vector $(\Delta t/(2m),1-\gamma\Delta t/2)$. This large axis
corresponds to the contribution of $\sigma\Delta W= \sigma\Wien{t}{t+\Delta
  t}$.  The narrow axis corresponds to the $\sigma\Delta g$ contribution,
giving the increment $(-\sigma\Delta g/m,\sigma\gamma\Delta g)$
(Fig.~\ref{fig:riceGrain}, left). By comparison, the Ricci and Ciccotti
algorithm, in this representation, reduces to a thin line with no width
(Fig.~\ref{fig:riceGrain}, right).  No matter how the algorithm is designed,
it requires two independent random numbers to sample the probability
distribution function corresponding to the Fokker-Planck propagator. In the
same way, when dealing with $3N$ degrees of freedom, the sampling of the
conditional probability associated to the propagator requires $6N$ random
numbers.

The relevance of the random numbers $\Delta g$ depends on whether one is
interested in a strong or a weak convergence of the algorithm. Were the random
force $\psi(t)$ known, it would make sense to estimate the difference between
the numerical solution and the exact solution, and in that case, the Ermak
scheme would be superior to the stochastic Verlet scheme, making it a true
strong order 1.5 algorithm. However, as far as numerical simulations are
concerned, we are interested in averaging the observables over many
realizations of the noise $\psi(t)$ (either by running the simulation long
enough, or by repeating many times the simulation process) and this
corresponds to the weak convergence concept, where only the difference between
the averaged simulated quantities and the averaged exact quantities matters.
As the variance of $\Delta g$ is $\Delta t^3$, as the variance of 
$\Delta W$ is only known up to order $\Delta t^2$, and as $\Delta g$ and
$\Delta W$ are statistically independent, the average contribution of
$\Delta g$ is at best of the same order of magnitude as the uncertainty
affecting $\Delta W$.  Not providing real ``information'', the generation of
the numbers $\Delta g$ just makes the Ermak procedure a slightly better
sampling of the underlying stochastic dynamics than the other algorithms
based on a single random number. However, the difference is not
quantitative: both schemes are of weak order two. 

As far as the algorithms based on the Fokker-Planck approach are concerned,
such as the one of Serrano~\textit{et al.}~\cite{FabSerEspCov2006,
  SerFabEspCov2006}, discussed in the DPD section below, the resulting
algorithms are also weak order two.  This comes from the fact that the
Fokker-Planck approach assumes an implicit average of the random
noise, when writing the Fokker-Planck evolution operator.


\section{Langevin dynamics of fluids}
\label{sec:LangevinDynamics}

We now specialize the discussion to the Langevin dynamics of a fluid, and
consider $N$ particles with usual pairwise interactions. The Liouvillian
reads: 
\begin{equation}
\Liou(t) = \sum_{i=1}^N \sum_{\alpha=1}^{3} \left[ 
\frac{\bp_{i\alpha}}{m}\frac{\partial}{\partial \br_{i\alpha}} 
+ \bigg(\bf_{i\alpha}(\{\br\})-\gamma \bp_{i\alpha}
+\sigma\bpsi_{i\alpha}(t)\bigg) 
\frac{\partial}{\partial \bp_{i\alpha}} \right].
\end{equation}
where $\alpha$ refers to the space index. As above, the time ordered
exponential is replaced by a product~:
\begin{eqnarray}
\mathrm{T}\exp\left(-\int_{t}^{t+\Delta t} \Liou(s) \dd s\right) 
&=& \exp\left(-\int_{t}^{t+\Delta t} \Liou(s) \dd s\right)
\exp\left(-\frac{1}{2}\int_{t}^{t+\Delta t} \dd s
\int_{t}^{s}\dd u\, [\Liou(u),\Liou(s)] \right);
\label{eq:RCsimplify}\nonumber\\
&=& \exp(-A-B)\cdot\exp C,
\end{eqnarray}
with 
\begin{eqnarray}
A &=& \int_t^{t+\Delta t} \dd s \sum_{i=1}^N \sum_{\alpha=1}^{3}  
\frac{\bp_{i\alpha}}{m}\frac{\partial}{\partial \br_{i\alpha}} 
= \Delta t \sum_{i=1}^N \sum_{\alpha=1}^{3}  
\frac{\bp_{i\alpha}}{m}\frac{\partial}{\partial \br_{i\alpha}};\\
B &=& \int_t^{t+\Delta t} \dd s \sum_{i=1}^N \sum_{\alpha=1}^{3} 
\left[ \bf_{i\alpha}(\{\br_i\})-\gamma \bp_{i\alpha}
+\sigma\,\bpsi_{i\alpha}(s)\right] \frac{\partial}{\partial \bp_{i\alpha}};\\
&=& \sum_{i=1}^N \sum_{\alpha=1}^{3} \bigg\lbrace \Delta t \left[
\bf_{i\alpha}(\{\br_i\}) -\gamma \bp_{i\alpha}(\br)\right]
+\sigma\Wieni{t}{t+\Delta t}{i\alpha}\bigg\rbrace
\frac{\partial}{\partial\bp_{i\alpha}}.
\end{eqnarray}
Using the BCH formula, we obtain two factorizations~: 
\begin{align}
\exp(-A-B)\cdot\exp(C) &= \exp\Big(-\frac{A}{2}\Big) \cdot
\exp(-B)\cdot\exp\Big(-\frac{A}{2}\Big)\cdot\exp(C);\label{eq:splitting13}\\
 &= \exp\Big(-\frac{B}{2}\Big)\cdot\exp(-A)\cdot
\exp\Big(-\frac{B}{2}\Big)\cdot\exp(C).\label{eq:splitting14}
\end{align}
If we forget the commutator term $C$, we obtain exactly the splitting proposed
by Ricci and Ciccotti, but with the opposite sign convention. The commutator
term is~:
\begin{align}
C &=\frac{1}{2}\int_{t}^{t+\Delta t} \dd s \int_{t}^{\displaystyle
    s}\dd u [\Liou(s),\Liou(u)];\notag\\
 &= \sigma \sum_{i=1}^N \sum_{\alpha=1}^{3} 
\bigg[ \frac{\Delta t}{2}\Wieni{t}{t+\Delta t}{i\alpha}
\notag\\
 &\hspace{3cm} -\int_t^{t+\Delta t} \dd s\,
  \Wieni{t}{s}{i\alpha}\bigg] 
  \bigg(\frac{1}{m}\frac{\partial}{\partial \br_{i\alpha}} 
  -\gamma\frac{\partial}{\partial \bp_{i\alpha}}\bigg).
  \label{eq:LangevinTimeCommutator}
\end{align}
The term $\exp(C)$ adds an extra contribution of order $\Delta t^{3/2}$ to the
random Gaussian increments. For completeness, we give below the correct
discretization schemes associated to the two splitting~(\ref{eq:splitting13})
and~(\ref{eq:splitting14}). A detailed calculation is provided in
appendix~(\ref{app:vv}).  The splitting~(\ref{eq:splitting13}) leads to
\begin{align}
\bp_{i\alpha}(t+\Delta t) &= \bp_{i\alpha}(t)\left(1-\gamma\Delta t
  +\frac{\gamma^2\Delta  t^2}{2}\right) 
  +\bf_{i\alpha}\left(\br(t)+\bp(t)\frac{\Delta
  t}{2m}\right)\left(1-\frac{\gamma\Delta t}{2}\right)\Delta t\nonumber\\ 
  &\hspace{1cm}+\sigma\Wieni{t}{t+\Delta t}{i\alpha}
  -\sigma\gamma\int_t^{t+\Delta t}\dd s\,
  \Wieni{t}{s}{i\alpha};\label{eq:splittingABAP}\\  
\br_{i\alpha}(t+\Delta t) &= 
  \br_{i\alpha}(t)+ \frac{\Delta t}{m}\bp_{i\alpha}(t)
  \left(1-\frac{\gamma \Delta t}{2}\right) 
  +\bf_{i\alpha}(\br(t))\frac{\Delta t^2}{2m}
\nonumber\\
  &\hspace{1cm}+\frac{\sigma}{m}\int_t^{t+\Delta t}\dd s\,
  \Wieni{t}{s}{i\alpha}.
\label{eq:splittingABAR}
%
\end{align}
The random Gaussian increments $\Delta \bp^E$, $\Delta \br^E$ appearing 
in~(\ref{eq:splittingABAP}) and~(\ref{eq:splittingABAR}) can themselves be
split in two parts $\Delta \bp^E=\Delta \bp^R+\Delta \bp^C$, and 
$\Delta \br^E=\Delta \br^R+\Delta \br^C$.
\begin{eqnarray}
\Delta\br^E_{i\alpha} &=& 
 \frac{\sigma}{m}\int_t^{t+\Delta t}\dd s\,\Wieni{t}{s}{i\alpha};\\
\Delta\br^C_{i\alpha} &=& -\frac{\sigma}{m}\left\lbrace\frac{\Delta t}{2} 
  \Wieni{t}{t+\Delta t}{i\alpha}
  -\int_t^{t+\Delta t}\dd s\, \Wieni{t}{s}{i\alpha}
  \right\rbrace;\\
\Delta\br^R_{i\alpha} &=& \frac{\sigma}{m}\left\lbrace\frac{\Delta t}{2} 
  \Wieni{t}{t+\Delta t}{i\alpha}\right\rbrace;\\
\Delta\bp^E_{i\alpha} &=& \sigma\Wieni{t}{t+\Delta t}{i\alpha}
  -\sigma\gamma\int_t^{t+\Delta t}\dd s\,\Wieni{t}{s}{i\alpha};\\
\Delta\bp^C_{i\alpha} &=& \sigma\gamma\left\lbrace\frac{\Delta t}{2}
  \Wieni{t}{t+\Delta t}{i\alpha}
  -\int_t^{t+\Delta t}\dd s\,\Wieni{t}{s}{i\alpha}\right\rbrace;\\
\Delta\bp^R_{i\alpha} & = & \sigma\left(1-\frac{\gamma\Delta t}{2}\right)
 \Wieni{t}{t+\Delta t}{i\alpha}.
\end{eqnarray}
Clearly, $\Delta \bp^C$ and $\Delta \br^C$ come from the time
commutator~(\ref{eq:LangevinTimeCommutator}). It is safe to replace the
complex random increments $\Delta \bp^E$ and $\Delta \br^E$ by their simpler
counterpart $\Delta \bp^R$ and $\Delta \br^R$ without altering the weak order
two nature of the algorithm.

The splitting~(\ref{eq:splitting14}) leads to another expression:
\begin{eqnarray}
\bp_{i\alpha}(t+\Delta t) &=& \bp_{i\alpha}(t)\left(1-\gamma\Delta t
  +\frac{\gamma^2\Delta t^2}{2}\right) 
  +\frac{\Delta t}{2}\bigg(\bf_{i\alpha}(\br(t))
  +\bf_{i\alpha}(\br(t+\Delta t))\bigg)  
  -\gamma \frac{\Delta t^2}{2} \bf_{i\alpha}(\br(t))\nonumber\\
& & + \sigma \Wieni{t}{t+\Delta t}{i\alpha}
-\gamma\sigma\int_t^{t+\Delta t}\dd s\,\Wieni{t}{s}{i\alpha} ;
\label{eq:splittingBABP}\\ 
\br_{i\alpha}(t+\Delta t) &=& \br_{i\alpha}(t) 
 +\frac{\Delta t}{m}\bp_{i\alpha}(t) 
   \left(1-\gamma\frac{\Delta t}{2}\right)  
 +\bf_{i\alpha}(\br(t))\frac{\Delta t^2}{2m}\notag\\
& &  +\frac{\sigma}{m}\int_t^{t+\Delta t}\dd s\,\Wieni{t}{s}{i\alpha}.
 \label{eq:splittingBABR} 
\end{eqnarray}
The above algorithm is equivalent to (\ref{eq:basicErmakVerlet:r}),
(\ref{eq:basicErmakVerlet:p}) and (\ref{eq:CorrelationErmakVerlet}). The
random Gaussian increments are obviously identical at the order $\Delta t^2$
considered here.  Equations~(\ref{eq:splittingABAP}),
(\ref{eq:splittingABAR}), (\ref{eq:splittingBABP}) and
(\ref{eq:splittingBABR}) are the main result of this section.

We believe that the absence of $\Delta\br^C$ and $\Delta\bp^C$ in
Ref.~\cite{RicCic2003} is due to a time-ordered exponential hastily replaced
by an ordinary exponential. These terms should have been present, and there
absence can be traced back to the forward operator $\mathcal{J}$ appearing in
the Susuki formula $\exp(\Delta t\mathcal{J})$, which, in the presence of the
random noise, must acquire some kind of time-dependence, such as $\exp(\Delta
t \mathcal{J}(t))$. The derivation of the algorithms of Ref.~\cite{RicCic2003}
is incomplete, with a misprint in the
result~(\ref{eq:RicciCiccotti:r}),(\ref{eq:RicciCiccotti:p}).


\section{Dissipative particle dynamics}
\label{sec:DissipativeParticleDynamics}

\subsection{Derivation of two original DPD algorithms}
We now apply the same strategy to the dissipative particle dynamics~(DPD)
introduced in Ref.~\cite{HooKoe1992,HooKoe1993} and discussed in
Ref.~\cite{EspWar1995, Espanol1995}. The phase space dynamics of the DPD is
given by a set of stochastic equations:
\begin{eqnarray}
\dot{\br}_i &=& \frac{\bp_i}{m}; \\
\dot{\bp}_i &=& \bf_i(\{\br\}) +\sum_{j,(j\neq i)} -\frac{\gammat}{m}
w(r_{ij})^2\be_{ij}\cdot(\bp_i-\bp_j)\,\be_{ij} +\sum_{j,(j\neq i)} 
\sigma w(r_{ij})\be_{ij} \psi_{ij}(t).
\label{eq:SDEdynamics}
\end{eqnarray}
Each particle $i$ exchanges randomly some momentum with a selected set of
neighbors $j$. This exchange dynamics depends on an arbitrary short range
function $w(r_{ij})$, and acts along the separation $\br_j-\br_i$, where
$\br_{ij}$ stands for $\br_j-\br_i$, $r_{ij}$ for $||\br_{ij}||$ and the unit
vector $\be_{ij}$ for $\br_{ij}/r_{ij}$.  To each pair of particles is
associated a random noise $\psi_{ij}(t)\dd t$, even though only a fraction of
the pairs are truly interacting together at a time, due to the finite range
of $w(r)$. The prefactor $\sigma$ is equal to $\sqrt{2\gammat k_B T}$ in
agreement with the fluctuation dissipation theorem. Note that here, the
dissipation coefficient $\gammat$ is defined in such a way that the product of
$\gammat$ and a velocity is a force. In order to stay consistent with the
previous section, we introduce also a friction coefficient $\gammal = m
\gammat$, homogeneous to a frequency, and such as $\sigma=\sqrt{2m\gammal
  k_BT}$.
These equations can be rewritten with $3N$-dimensional vectors and matrices: 
$\br= \{\br_i\}$, $\bp= \{\bp_i\}$, $\partial_{\bp} =
\{\partial/\partial_{\bp_i}\}$, $\partial_{\br} =
\{\partial/\partial_{\br_i}\}$,
\begin{eqnarray}
\bR_{i\alpha}(\br;t) &=& \sigma\left\lbrace \sum_{j,(j\neq i)}
\frac{w(r_{ij})}{r_{ij}}(\br_i-\br_j)_{\alpha}\,\psi_{ij}(t)\right\rbrace;\\
\bd_{i\alpha}(\br,\bp) &=& -\gammal\left\lbrace \sum_{j,(j\neq i)}
\frac{w(r_{ij})^2}{r_{ij}^2}[(\br_i-\br_j)\cdot(\bp_i-\bp_j)] 
(\br_i-\br_j)_{\alpha} \right\rbrace.
\end{eqnarray}
The dissipative term $\bd(\br,\bp)$ can be expressed as the product of a space
dependent $3N\times 3N$ matrix $\bGamma(\br)$ and the $3N$-vector momentum
$\bp$. If $\alpha,\beta$ refer to the space indices, then the components of
the matrix $\bGamma(\br)$ read:
\begin{eqnarray}
\mathbf{\Gamma(\br)}_{i\alpha,j\beta} &=& \gammal \sum_{k,(k\neq i)}
\frac{w^2(r_{ik})}{r_{ik}^2}
(r_{i\alpha}-r_{k\alpha})(r_{i\beta}-r_{k\beta})(\delta_{ij}-\delta_{kj}); \\
\bd(\br,\bp) &=& -\bGamma(\br) \cdot \bp.
\end{eqnarray}  
It can be checked that the correlations of the noise obey 
\begin{eqnarray}
\mean{\psi_{ij}(t)} &=& 0;\notag\\
\mean{\psi_{ij}(t)\psi_{kl}(t')} &=& 
(\delta_{ik}\delta_{jl}+\delta_{il}\delta_{jk})\delta(t-t');\\
\mean{\bR(\br;t)_{i\alpha}} &=& 0;\notag\\
\mean{\bR(\br;t)_{i\alpha}\bR(\br;t')_{j\beta}} &=& 2\gamma m k_B T\delta
(t-t') \bGamma(\br)_{i\alpha,j\beta}.
\end{eqnarray}
To the s.d.e.~(\ref{eq:SDEdynamics}) we associate  the Liouvillian 
\begin{equation}
\Liou(t) =\left( \frac{\bp}{m}\cdot\partial_{\br} 
 +\bf(\br)\cdot\partial_{\bp} 
 +\bd(\br,\bp)\cdot\partial_{\bp}\right) 
 +\bR(\br;t)\cdot \partial_{\bp}.
\end{equation} 
The main difference, compared with the ordinary Langevin dynamics, is that the
prefactor of the noise depends explicitely on $\br$. It is known, however,
that this multiplicative noise is not dependent of the It\^{o} or the
Stratonovitch interpretation of stochastic dynamics~\cite{EspWar1995}.
Regarding the simple exponential part $\exp(-\int\Liou(s)\dd s)$, we define, as
usual  
\begin{eqnarray}
A &=& \Delta t\,\frac{\bp}{m}\cdot\partial_{\br};\\
B &=& \Delta t\,[\bf(\br)+\bd(\br,\bp)]\cdot\partial_{\bp}
+\bigg[\int_{t}^{t+\Delta t} \bR(\br;s)\dd s\bigg] 
\cdot\partial_{\bp},
\label{eq:definitionAB:dpd}
\end{eqnarray}
and the $3N$ vector $\bK(\br;t)$:
\begin{equation}
\bK(\br;t) = \bf(\br)+\frac{1}{\Delta t}
\bigg[\int_t^{t+\Delta t}\bR(\br;s)\dd s\bigg].
\end{equation}
$A$ scales like $\Delta t$, $B$ scales like $\sqrt{\Delta t}$, $K(\br;t)$
scales like $\Delta t^{-1/2}$. One possible splitting is
\begin{multline}
\displaystyle\mathrm{T}
\exp\left(-\int_t^{t+\Delta t} \Liou(s)\dd s\right) = 
\displaystyle\exp\Big(-\frac{B}{2}\Big)\cdot\exp\Big(-A\Big)\cdot
\exp\Big(-\frac{B}{2}\Big)\\ 
\displaystyle\cdot\exp\bigg(\frac{1}{24} \lbrack B,[A,B]\rbrack
-\frac{1}{2}\int_t^{t+\Delta t}\dd s\,\int_{t}^{s}\dd u\,
 [\Liou(u),\Liou(s)]\\ 
-\frac{1}{3}\int_t^{t+\Delta t}\dd s\,\int_{t}^{s}\dd u\,\int_{t}^{s}\dd v\,
[\Liou(v),[\Liou(u),\Liou (s)]]\bigg),
\label{eq:splittingDPDBAB}
\end{multline}
and the other is
\begin{multline}
\displaystyle\mathrm{T}
\exp\left(-\int_t^{t+\Delta t} \Liou(s)\dd s\right) = 
\exp\Big(-\frac{A}{2}\Big)\cdot\exp\Big(-B\Big)\cdot
\exp\Big(-\frac{A}{2}\Big)\\ 
\displaystyle
\cdot\exp\bigg(-\frac{1}{12} \lbrack B,[A,B]\rbrack
-\frac{1}{2}\int_t^{t+\Delta t}\dd s\,\int_{t}^{s}\dd u\,
[\Liou(u),\Liou (s)]\\
-\frac{1}{3}\int_t^{t+\Delta t}\dd s\,\int_{t}^{s}\dd u\,\int_{t}^{s}\dd v\,
[\Liou(v),[\Liou(u),\Liou (s)]]\bigg).
\label{eq:splittingDPDABA}
\end{multline}
The splitting gives the largest contribution, while the commutators give
smaller correcting terms. The first splitting $\exp(-A/2)\exp(-B)\exp(-A/2)$
gives after neglecting terms smaller than $\Delta t^2$ (algorithm ``ABA''),
\begin{eqnarray} 
\br(t+\Delta t) &=& \br(t) + \frac{\Delta t}{2m} 
\bigg[\bp(t)+\bp(t+\Delta t)\bigg];
\label{eq:splittingDPDABA:R}\\
\bp(t+\Delta t) &=& \left[\bm{1}-\Delta t\,\bGamma\bigg(\br(t)
 +\frac{\Delta t}{2m}\bp(t)\bigg) +\frac{\Delta t^2}{2}\bGamma^2\bigg(\br(t)
 +\frac{\Delta t}{2m}\bp(t)\bigg)\right]\cdot \bp(t) \notag\\
& &  + \Delta t\left[\bm{1}-\frac{\Delta t}{2}\bGamma\bigg(\br(t) 
 +\frac{\Delta t}{2m}\bp(t)\bigg)\right]
\cdot\bK\bigg(\br(t)+\frac{\Delta t}{2m}\bp(t);t\bigg),
\label{eq:splittingDPDABA:P}
\end{eqnarray}
while $\exp(-B/2)\exp(-A)\exp(-B/2)$ gives
\begin{eqnarray} 
\br(t+\Delta t) &=& \br(t) +\frac{\Delta t}{m}\bp(t)
-\frac{\Delta t^2}{2m}\bGamma(\br(t))\cdot\bp(t)
+\frac{\Delta t^2}{2m} \bK(\br(t);t);
\label{eq:splittingDPDBAB:R}\\ 
\bp(t+\Delta t) &=& \bp(t) -\frac{\Delta t}{2}
\bigg[\bGamma(\br(t))+\bGamma(\br(t+\Delta t))\bigg]\cdot\bp(t) 
+\frac{\Delta t^2}{2}
\bGamma(\br(t))^2\cdot\bp(t) \nonumber\\
& & 
+\frac{\Delta t}{2}\bigg[\bK(r(t);t) + \bK(r(t+\Delta t);t)\bigg] 
-\frac{\Delta t^2}{2} \bGamma(\br(t))\cdot \bK(\br(t);t).
\label{eq:splittingDPDBAB:P}
\end{eqnarray}
We now turn to the double and triple commutator corrections.  The correction
$\exp(-1/2\iint\dd u\,\dd s\,[\Liou(u),\Liou(s)])$ gives terms proportional to
\hbox{$\int_t^{t+\Delta t} (s-\Delta t/2) \psi_{ij}(s)\dd s$}, which have been
shown to be uncorrelated to the $\dpdWien{t}{t+\Delta t}{ij}$, of vanishing
mean value and of variance $\Delta t^3$. These terms are the pendant of the
corrections in the Ermak algorithm, and can be neglected in practice.

After some calculations exposed in appendix~(\ref{app:dpd}), it appears that 
together the correction of the triple commutators is either vanishing 
for the splitting~(\ref{eq:splittingDPDABA}), or equal to
\begin{equation}
\Delta p_{i\alpha}=
-\frac{\sigma^2\Delta t^2}{4m} 
\sum_{j,(j\neq i)}\frac{\partial}{\partial r_{i\alpha}}
\Big(w(r_{ij})^2\Big),
\label{eq:correctionTermDPDFinal}
\end{equation}
for the splitting~(\ref{eq:splittingDPDBAB}). A popular choice consists of
using a triangular shape for $w$. 
\begin{equation}
\left\lbrace
\begin{array}{cccc}
w(r) &=& 1-r/r_c, &\;\mathrm{if}\; r<r_c;\\
w(r) &=& 0, &\;\mathrm{if}\; r\geq r_c.\\
\end{array}
\right.
\end{equation}
This choice ensures that $w^2$ is differentiable everywhere but in $r=0$, and
the corrective term is well defined. 
\begin{equation}
\frac {\partial}{\partial r_{i\alpha}} \frac{(r_c-r_{ij})^2}{r_c^2} 
= -2\frac{r_c-r_{ij}}{r_c^2 r_{ij}}(\br_{i\alpha}-\br_{j\alpha})
= 2\frac{r_{ij}-r_c}{r_c^2}\be_{ij,\alpha}. 
\end{equation}
This suggests the final form of our DPD algorithm, corresponding to the
splitting~(\ref{eq:splittingDPDBAB:R}) and (\ref{eq:splittingDPDBAB:P})
(algorithm ``BAB'').
\begin{eqnarray} 
\br(t+\Delta t) &=& \br(t)+\frac{\bp(t)}{m}\Delta t
 -\frac{\Delta t}{2m}\bGamma(\br(t))\cdot\bp(t)\Delta t 
 +\frac{\Delta t^2}{2m}\bK(\br(t);t)
\label{eq:algoDPDBAB:R}\\  
\bp(t+\Delta t) &=& \bp(t) -\frac{\Delta t}{2}
  \bigg[\bGamma(\br(t))+\bGamma(\br(t+\Delta t))\bigg]\cdot\bp(t) 
  +\frac{\Delta t^2}{2}\bGamma(\br(t))^2\cdot\bp(t) \notag\\
& &  +\frac{\Delta t}{2}\bigg[\bK(\br(t);t)+\bK(\br(t+\Delta t);t)\bigg]
  -\frac{\Delta t^2}{2} \bGamma(\br(t))\cdot \bK(\br(t);t)\notag\\
& &  -\frac{\sigma^2\Delta t^2}{4m} \mathbf{g}_w\cdot\br,
\label{eq:algoDPDBAB:P}
\end{eqnarray}
where the last term $\mathbf{g}_w\cdot\br$ requires a $3N\times3N$ operator
matrix  
\begin{equation}
(\mathbf{g}_w)_{i\alpha,j\beta} = \frac{1}{3}\frac{\partial}{\partial
  r_{i\alpha}}w(r_{ij})^2  \frac{\partial}{\partial r_{j\beta}},
\label{eq:BABCorrectiveTerm}
\end{equation}
and
\begin{equation}
\mathbf{g}_w\cdot\br = \sum_{j\neq i,\alpha}\frac{\partial}{\partial
  r_{i\alpha}} \Big(w(r_{ij})^2\Big).
\end{equation}

Formula~(\ref{eq:splittingDPDABA:R}) and~(\ref{eq:splittingDPDABA:P}), on the
one hand, and~(\ref{eq:algoDPDBAB:R}), (\ref{eq:algoDPDBAB:P}) and
(\ref{eq:BABCorrectiveTerm}) on the other hand, are the main results of this
section. In the following section, we give more details on how to implement
these formulas in practice.

\subsection{Comparison with other existing algorithms}

The BAB algorithm resembles the DPD-velocity Verlet (DPD-VV)
algorithm~\cite{GrootWarren1997}. Both use an estimate
$[\bf(\br(t))+\bf(\br(t+\Delta t))]/2$ for the forces, and a similar
expression for the position dependent dissipative forces. Positions are
updated in an identical way, positions and velocities are updated in the same
consecutive order. However, a noticeable difference concerns the use of the
random numbers~: in our case, we use a single random number
$\Wien{t}{t+\Delta t}$, while DPD-VV use a combination of $\Wien{t-\Delta
  t/2}{t}$ and $\Wien{t}{t+\Delta t/2}$. Moreover some terms differ at the
order $\Delta t^2$.

Starting from a Fokker-Planck formulation of the stochastic dynamics, the
approach by Serrano \textit{et al.}~\cite{FabSerEspCov2006,SerFabEspCov2006}
gives directly a weak order two algorithm, and by-passes all the complications
related to the time-dependent double and triple commutators in the Magnus
expansion.  Although we do not present a demonstration here, it is also
possible to derive their algorithm starting from our
splitting~(\ref{eq:splittingDPDABA}). One of the steps in the derivation
consists in splitting further the exponential of the operator~$B$ defined
in~(\ref{eq:definitionAB:dpd}) into a product over all pairs of particles.

In an improvement over DPD-VV, Pagonabarraga \textit{et
  al.}~\cite{2006_Pagonabarraga_Frenkel} proposed a self-consistent procedure
for computing the velocity dependent friction term.  This makes the so called
self-consistent DPD-VV (SC-DPD-VV) a reversible algorithm, for which, in the
absence of potential forces, the stochastic trajectories can be traced back
upon momentum reversal, according to the detailed balance probabilities.

We did not find mention of the reversibility properties of the algorithms of
Shardlow~S2 and Serrano \textit{et al.}, but we believe that, much as the
Ermak procedure for Langevin dynamics, the algorithm of Serrano \textit{et
  al.}  stands a good chance to be reversible, because the momentum exchange
stochastic dynamics is treated without approximation. The algorithm S2 of
Shardlow, which almost reduces to the former in the absence of potential
forces, can only be approximately reversible, while retaining the weak order
two convergence property.

As far as our algorithms are concerned, the stochastic evolution of the
momenta $\bp(t)$, with fixed positions $\br(t)$, corresponds to a
multidimensional Ornstein-Uhlenbeck process, as illustrated by
eq.~(\ref{eq:exampleReversible}) of appendix~\ref{app:dpd}.  Reversibility is
lost when approximations such as $\exp(-\bGamma\Delta t)\simeq 1
-\bGamma\Delta t + \bGamma^2 \Delta t^2/2$ are introduced. As keeping
exponentials of matrices is not algorithmically feasible, the final form of
the algorithms BAB and ABA are only approximately reversible, up to the
expected order of convergence $\Delta t^2$.


\newcommand{\ga}{\gamma}
\section{Numerical tests and discussion}
\label{sec:NumericalDiscussion}

\subsection{Ermak versus stochastic Verlet algorithms}

We noticed a fundamental difference between the Ermak and Ricci-Ciccotti (or
stochastic Verlet) algorithms, the former being in principle slightly more
accurate in simulating the Langevin dynamics. It is interesting to probe
numerically whether such an assertion is really detectable. To this end, we
considered the simple Ornstein-Uhlenbeck process :
\begin{align}
  \ddot{x}+\dot{x}+x&=\psi(t),\\
\langle\psi(t)\psi(t')\rangle&=2\delta(t-t').
\end{align}
which is exactly solvable. In particular, we have
\begin{align}
\langle x(t)x(0)\rangle_\text{exact}=
e^{-t/2}\times\left(\cos(\sqrt{3}t/2)+\frac{1}{\sqrt{3}}\sin(\sqrt{3}t/2)\right).       
\end{align}
A way to check the performances of an algorithm is to measure how fast
and how accurately it reproduces such a correlator. Thus we define the
error associated with a numerical result as
\begin{align}
  \text{Error}=\frac{1}{\Tav}\int_0^{\Tav} \dd t \bigg[  
  \langle x(t)x(0)\rangle_\text{exact}-  
  \langle x(t)x(0)\rangle_\text{simulated}\bigg]^2
\end{align}
(more exactly the discrete version of this formula). $\Tav$ is a predefined
time kept constant. This error is a function of the random numbers used to
generate $\langle x(t)x(0)\rangle_\text{simulated}$, and can be considered as
a function of the simulation time $t_\text{sim}$ , as the average
$\langle\ldots\rangle$ is constructed by accumulation (running average). Were
the time step infinitely small, would such a curve decrease toward zero;
actually, it saturates (with fluctuations) at a higher value due to the
unavoidable discreteness of the simulations. Constructed with a single
simulation, this measure of the error as a function of the simulation length
is a noisy curve, and it is more relevant to consider the curve obtained by
averaging several such errors.

\begin{figure}[ht]
  \begin{center}
   \rotatebox{-90}{\resizebox{0.7\textheight}{!}{\includegraphics*{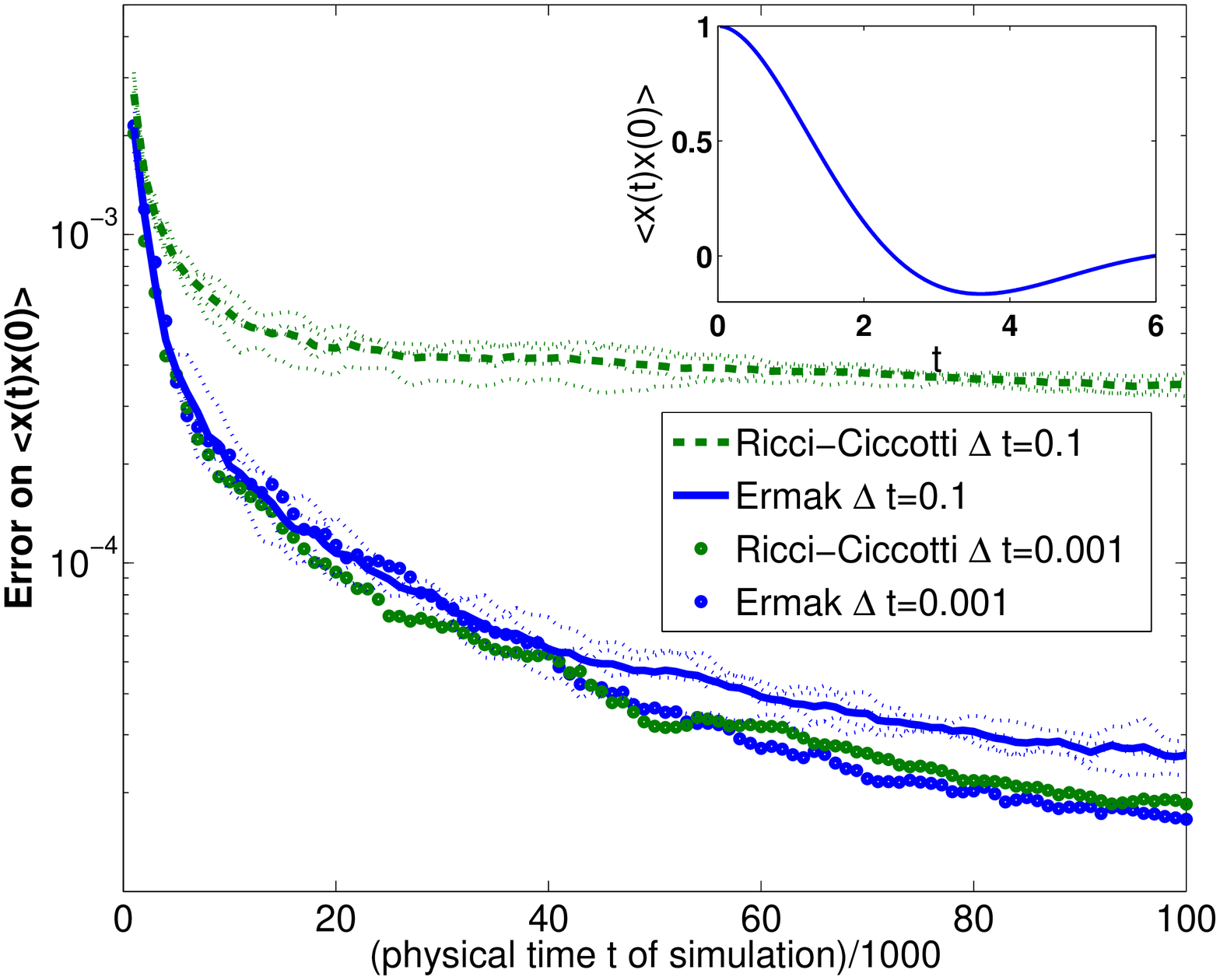}}}
  \caption{}
  \label{fig:errors_E_RC}
  \end{center}
\end{figure}

The result of such a procedure is plotted on Fig.~\ref{fig:errors_E_RC},
for eighty runs of $10^6$ time steps (full and dashed curves --- the light
dots show the curves for four subsets of twenty curves to show the
dispersion).

The chosen time $\Tav=6$ contains the essential part of $\langle
x(t)x(0)\rangle$ (see inset); the time step is large : $\Delta t=0.1$, large
enough to reveal in that situation the clear superiority of the Ermak
algorithm over the Ricci-Ciccotti one : for those large time steps, the Ermak
algorithm mimics a thermal noise which is less colored than Ricci-Ciccotti's,
and hence is a more faithful realization of Langevin dynamics. Obviously, when
reducing the time step, the spectrum of the noise is flatter and flatter for
both algorithms and the discrepancies disappear : this is shown on the curves
in large dots, obtained for a time step $\Delta t=0.001$ and $10^8$ simulation
points (the associated physical time is unchanged); they show that a
computational effort a hundred times greater than before leads to a rather
modest improvement as far as the Ermak algorithm is concerned.  But the Ermak
algorithm can be useful for situations where large time steps can be used,
without compromising a faithful sampling of the underlying Hamiltonian
dynamics. We think that these results suggest to always prefer a larger time
step with the Ermak algorithm to a smaller one with the lightest (in term of
computational cost) Ricci-Ciccotti scheme.

\subsection{Implementation of the ABA algorithm} 
Two algorithms of similar predicted accuracy are available.  In practice,
after having tried them both, we found that the algorithm
(\ref{eq:splittingDPDABA:R},\ref{eq:splittingDPDABA:P}) was by far the
simplest and the most efficient of the two. Therefore, we devote to this
scheme the largest part of our discussion, while briefly mentioning the other
at the end of this section. We chose to refer to it as the ``ABA'' algorithm,
by reference to the splitting from which it originates.

It reads explicitly:
\begin{center}
\begin{tabular}{cccl}
\hline\hline
(a) & $\br'_i$   & = & $\displaystyle\br_i(t)+\frac{\Delta t}{2m}\bp_i(t)$\\
(b) & $\bm{X}_i$ & = & $[\bm{\Gamma}(\br')\bp(t)]_i\Delta
t-\bm{K}_i(\br')\Delta t$\\
   & & = & $\gamma\Delta t\sum_{j,(j\neq i)} 
 \frac{w^2(r'_{ij})}{r_{ij}^{'2}}(\br'_{ij}.\bp_{ij})\br'_{ij}
 +\sigma\sqrt{\Delta t}\sum_{j,(j\neq i)} 
 \frac{w(r'_{ij})}{r'_{ij}}\br'_{ij}W_{ij}-\bm{f}_i(\br')\Delta t$\\
(c) & $\bm{Y}_i$ & = & $[\bm{\Gamma}(\br')\bm{X}(t)]_i$\\
& & = & $ \gamma\sum_{j,(j\neq i)} 
  \frac{w^2(r'_{ij})}{r_{ij}^{'2}}(\br'_{ij}.\bm{X}_{ij})\br'_{ij}$\\
(d) & $\bp_i(t+\Delta t)$ & = & $\displaystyle\bp_i(t)- \bm{X}_i+\frac{\Delta
  t}{2m} \bm{Y}_i$\\
(e) & $\br_i(t+\Delta t)$ & = & $\br'_i+\frac{\Delta t}{2}\bp_i(t+\Delta t)$,\\
\hline\hline
\end{tabular}
\end{center}
where we defined $\bm{x}_{ij}\equiv\bm{x}_i-\bm{x}_j$.  The $W_{ij}$ are
uncorrelated Gaussian random variables with zero mean and unit variance.  This
algorithm can be made even more efficient by merging the first and last steps
during the simulation (however, the computation of statistical quantities must
be performed compulsorily between these steps, otherwise a noticeable bias is
introduced). The computational effort per time step is slightly higher than
the Shardlow algorithm (eq.~(9) of Ref.~\cite{shardlow} hereafter termed
S1 algorithm according to the terminology used in~\cite{lesfinlandais}), as
here two computations implying the sums $\sum_j w(r_{ij})$ are required,
instead of just one for S1. This is the price to pay to have a second order
algorithm.  Actually we measured only an increase of ca. 5\% of the time
simulation with our algorithm with respect to S1 (to be efficient, the program
must maintain a subdivision of the simulation box and a related list of
occupation, or alternatively a neighbor list which already exists anyway for
MD, such that a loop $\sum_j w(r_{ij})$ is fast).

\subsection{Benchmarking with existing algorithms}

When a new algorithm is introduced, it is common practice to rate its
performance, and to compare it with the other existing algorithms. The usual
benchmark includes a calculation of the pair distribution function $g(r)$ of a
noninteracting fluid (ideal gas), a monitoring of the thermalization
properties, and a presentation of the self-diffusion data. The absence of a
non trivial but exactly solvable model makes that there is no stringent test
available for assessing the ability of an integrator to reproduce
faithfully the true DPD dynamics. Previous studies discovered that most DPD
algorithms do not predict correctly the flat shape of the $g(r)$
characterizing the ideal gas. Instead, some unphysical structure of
$g(r)$ near the origin is observed, betraying a dynamics with unsatisfactory
detailed balance properties.

We present data for both BAB and ABA algorithms, along with the Shardlow~S1
and DPD-VV algorithms (which are \textit{a priori} weak order one algorithm).
The scale and axis of our graphs make our data readily comparable with the one
presented by Nikunen~\textit{et al.}~\cite{lesfinlandais} or
Serrano~\textit{et al.}~\cite{SerFabEspCov2006}.  Fig.~\ref{fig:standardgr}
displays the $g(r)$ distribution, for the usual weight function $w(r) =
1-r/r_c$, and three increasing values of the time step $\Delta t=0.01$,
$\Delta = 0.05$, and $\Delta t=0.1$.  Our $g(r)$ is very satisfactory for
$\Delta t=0.01$, comparable with DPD-VV but worse than Shardlow~S1 for $\Delta
t=0.05$, and definitely worse than its competitors for $\Delta t=0.1$, a quite
large value by the way.  We also plot the BAB algorithm deprived of its
correcting term, for which the case $\Delta t=0.01$ clearly demonstrates that
this correcting term precisely contributes very much to the accuracy of the
scheme at small $\Delta t$, and supports firmly the validity of our
calculation.

Fig.~\ref{fig:standardT} summarizes the thermalization properties.  The left
panel of Fig.~\ref{fig:standardT} shows the simulated kinetic temperature
$\mean{T}$, compared with the value $T_{\rm thermostat}=1$ that the thermostat
attempts to enforce.  In the right panel of Fig.~\ref{fig:standardT}, we
present a semi-logarithmic plot of the difference between the actual simulated
temperature and its target value $|\mean{T}-T_{\rm thermostat}|$, for the same
three time steps $\Delta t=0.01$, 0.05 and 0.1. We draw from these data the
same conclusion as for the $g(r)$~: both algorithms ABA and BAB are found
inaccurate for $\Delta t=0.1$, they are comparable with DPD-VV for $\Delta
t=0.05$ while standing below Shardlow S1, and finally become as satisfactory
as the other algorithms for $\Delta t=0.01$. 

Fig.~\ref{fig:interactiongr} proves that our algorithms predicts correctly the
shape of the pair distribution in the presence of interaction, for the
``model~B'' introduced in Ref.~\cite{shardlow}, where an interparticle pair
potential equal to $aw^2(r)/2$; $a=25$ is present. Fig.~\ref{fig:diffusion}
summarizes the results obtained for the self-diffusion coefficient $D$, in the
presence of interactions between particles (``model~B''). In the absence of
any exact result for this non-trivial situation, we take the self-diffusion
value $D(\Delta t=0.01)$ obtained for the smallest time step as a reference,
and compare it with the values obtained for larger time step $\Delta t=0.05$
and $0.1$.  It is also possible to do the same study for an ideal gas, but we
found that all the tested algorithms (DPD-VV, Shardlow~S1 and ABA) lead
roughly to the same value. We believe that the data presented here are more
discriminant, with a combination of caging effect due to the interactions and
correlated diffusion due to the non-local dissipation term.

\subsection{The singularity of the weight functions $w(r)$}

The ideal gas has been widely used as a benchmark to test DPD integrators,
because of the notorious artefacts in $g(r)$ from which so many DPD algorithms
suffer. We claim that some attention must be paid in that case: for a choice
of a weight function with $w(r=0)\neq 0$, the DPD dynamics is singular
whenever two particles occupy the same position; therefore, the ideal gas (the
model \textit{par excellence} where superposition of particles can occur)
should only be used as a benchmark model for testing DPD algorithms for weight
functions such that $w(r)/r$ does not diverge when $r\rightarrow 0$.
Otherwise, what is measured is also the ability of the algorithm to deal with
this singularity, whereas the algorithms are always written under the
assumption of local differentiability of the functions defining the model.
This point is crucial: a high order numerical scheme may enhance the effect of
a singularity and perform worse than its lowest order competitors.

Fig.~\ref{fig:regulargr} demonstrates how efficient a regular weight function
is at removing the artefacts in the pair distribution of ideal gases. It shows
the $g(r)$ obtained for $w_1(r) = (r/r_c)\times(1-r/r_c)$, while the
thermalization properties of $w_1(r)$ are already shown in
Fig.~\ref{fig:standardT}. The weight function $w_1(r)$ vanishes for $r=0$,
precisely when a pair of particles is colliding, and preventing the sudden
reversal of the friction forces associated to this pair.
Clearly, the use of $w_1(r)$ for soft core particles generate no artefact in
$g(r)$, without reducing the thermalization efficiency of the DPD thermostat.
The comparison between the different algorithms turns now to the advantage of
our ABA scheme, with a flatter $g(r)$ than both DPD-VV and Shardlow S1 (right
panel of Fig.~\ref{fig:regulargr}).

\subsection{Discussion}
First, we would like to emphasize some shortcomings of the above tests,
focusing only on static correlators~: from the measurement of $g(r)$, it is
not possible to really check how faithful to the true DPD dynamics these
algorithms are; as an extreme illustration of this idea, a Ermak-Langevin
algorithm could be qualified as a good DPD algorithm, if only static
quantities are considered.  Actually, a genuinely discriminant test of DPD
algorithms can be made only in the spirit of the above comparison between
Ermak and Ricci-Ciccotti algorithms, that is using as a benchmark a dynamical
correlator.

A second question concerns the value of the damping coefficient $\gamma$,
which for the sake of the comparison with previously published data, was
fixed to a quite large value $\ga=4.5$, and for which an efficient
thermalization is achieved at the expense of a ``sluggish''dynamics. Usually,
second order algorithms are designed to match smoothly the underlying
Hamiltonian dynamics in the limit $\gamma\rightarrow 0$. Thus, for large
values of $\ga$, the putative force of our algorithms is not apparent,
compared with simpler order one schemes.

A related question arises~: provided that a numerical scheme samples
faithfully the canonical ensemble and conserves well the total momentum, is
that of any importance to stick to the actual DPD dynamics or not? The answer
is certainly yes if one is interested in making connections between numerical
results and theoretical developments. In this respect, our second order
algorithm is useful, as it matches better Hamiltonian and DPD
dynamics in the limit $\ga\rightarrow 0$, an interesting limit for those who
want the DPD only to thermally stabilize a long simulation without perturbing
its natural proper dynamics~\cite{dunweg}.

\begin{figure}[ht]
  \begin{center}
   \rotatebox{-90}{\resizebox{\textheight}{!}{\includegraphics*{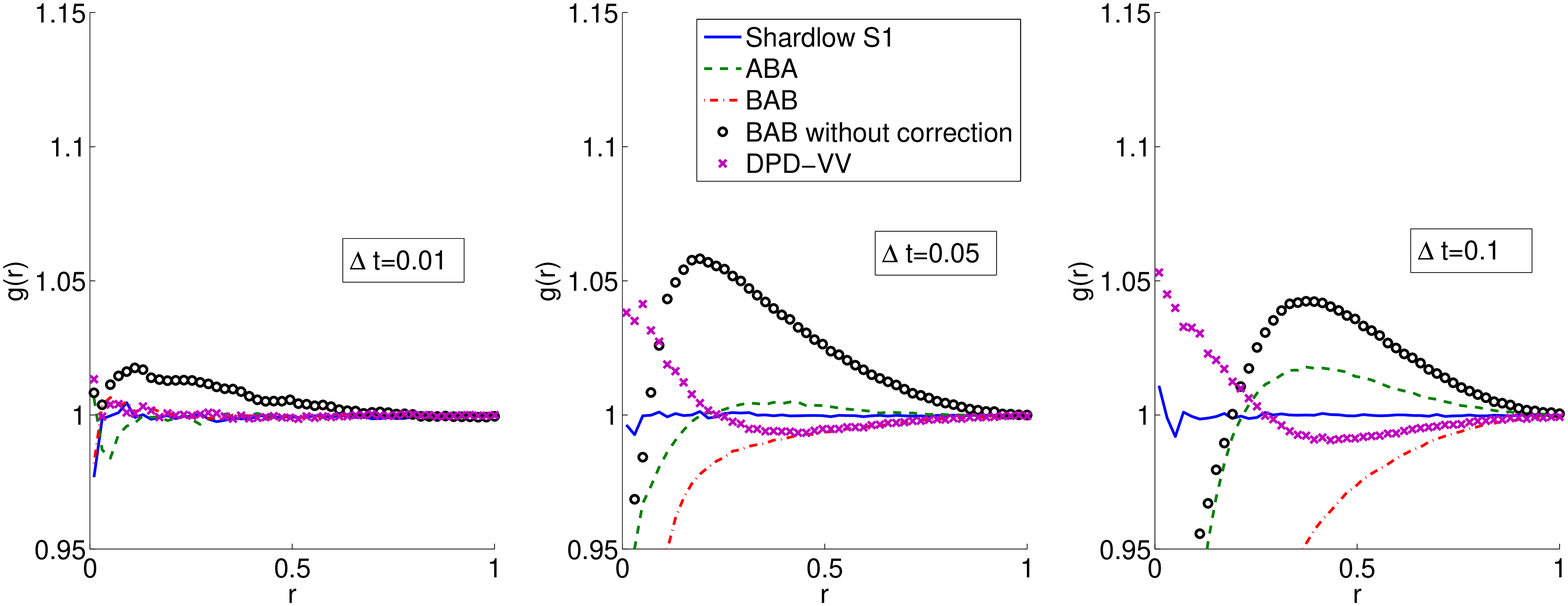}}}
  \caption{}
  \label{fig:standardgr}
  \end{center}
\end{figure}

\begin{figure}[ht]
  \begin{center}
  \rotatebox{-90}{\resizebox{\textheight}{!}{\includegraphics{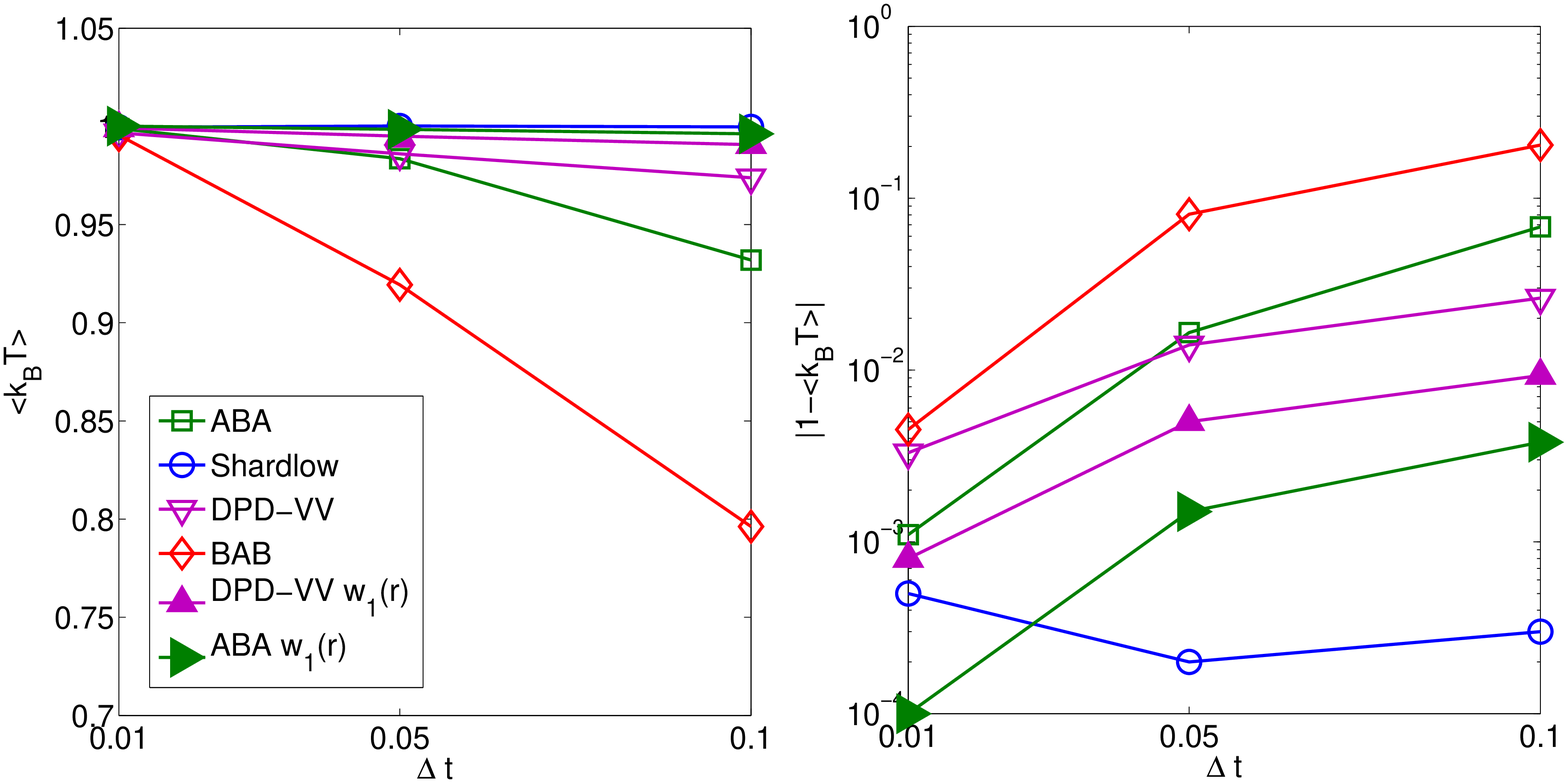}}}
  \caption{}
  \label{fig:standardT}
  \end{center}
\end{figure}

\begin{figure}[ht]
  \begin{center}
  \rotatebox{-90}{\resizebox{\textheight}{!}{\includegraphics*{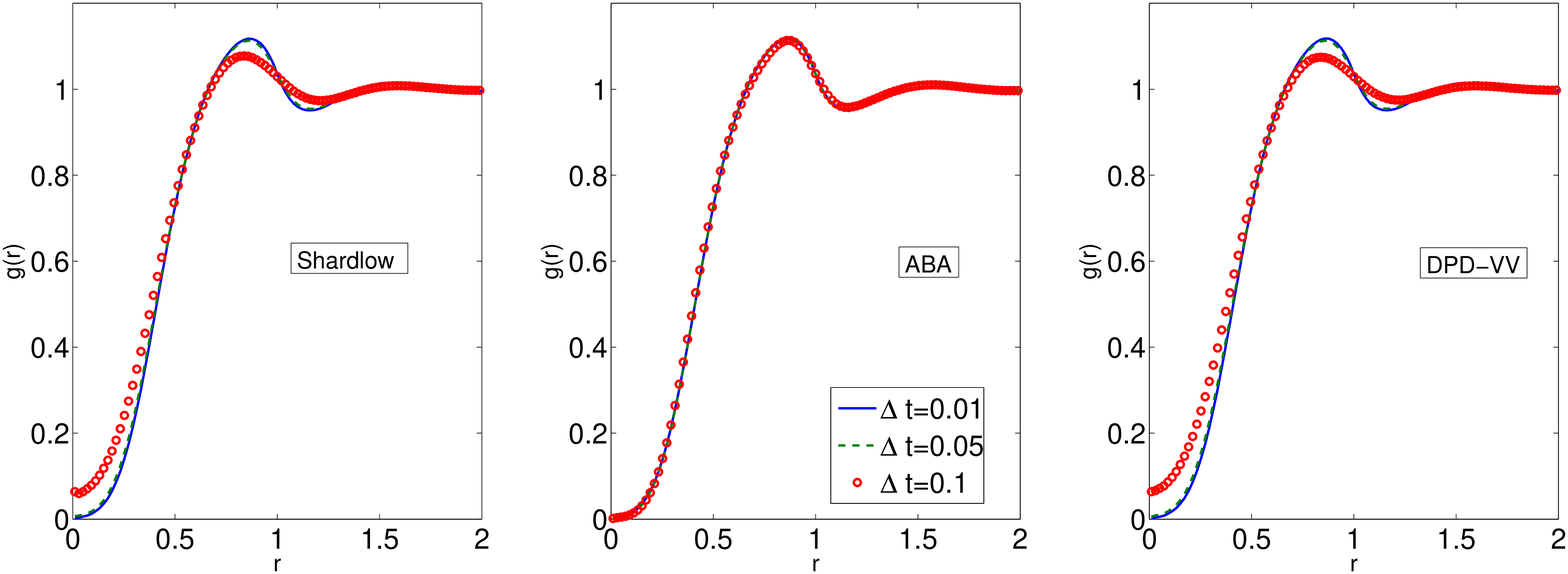}}}
  \caption{}
  \label{fig:interactiongr}
  \end{center}
\end{figure}

\begin{figure}[ht]
  \begin{center}
  \rotatebox{-90}{\resizebox{\textheight}{!}{\includegraphics*{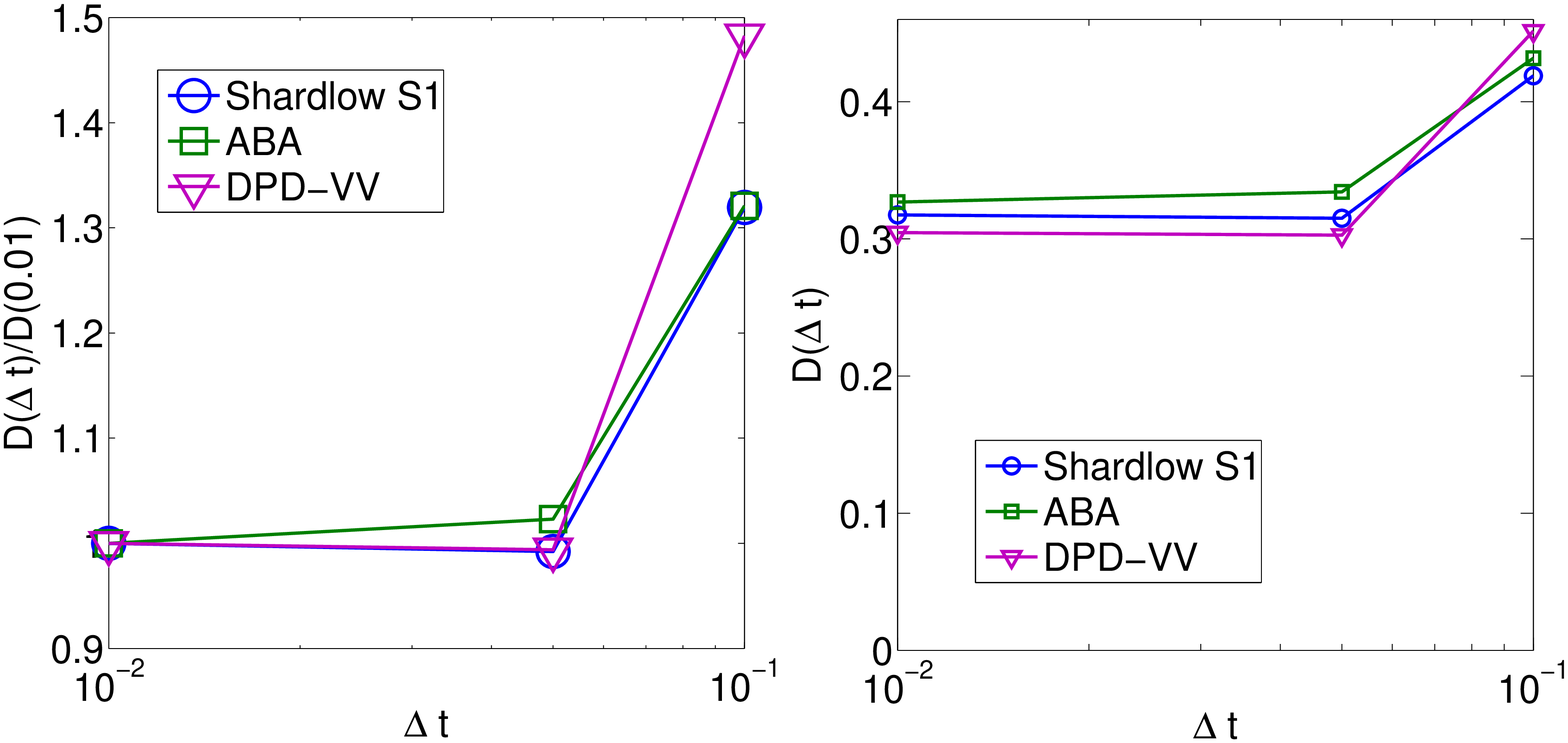}}}
  \caption{}
  \label{fig:diffusion}
  \end{center}
\end{figure}

\begin{figure}[h]
  \begin{center}
\rotatebox{-90}{\resizebox{1.1\textheight}{!}{\includegraphics*{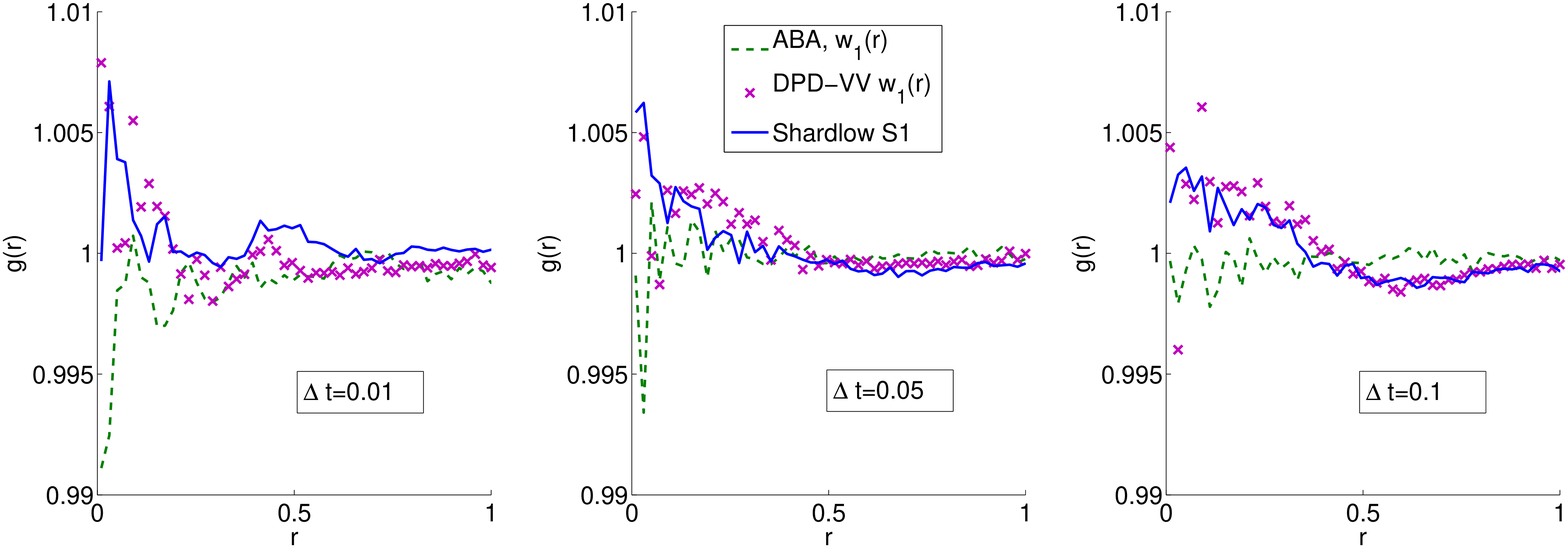}}} 
  \caption{}
  \label{fig:regulargr}
  \end{center}
\end{figure}

In our view, it is still an open question to understand why the algorithm of
Shardlow and of Ref.~\cite{SerFabEspCov2006} are so efficient in avoiding the
artefacts of $g(r)$ in Fig.~\ref{fig:standardgr}. We simply attribute this to
the sequential update of the velocities over all the pairs of particles,
rather than doing it all at once, as we propose. We do not, however, consider
Fig.~\ref{fig:standardgr} as a failure of our algorithms, but much more as a
warning message for the DPD users interested in soft core potentials,
concerning the inadequacy of the weight function $w(r)=1-r/r_c$ in that case.
Simulations with soft core particles should be done with a regular weight such
as $w_1(r)=(r/r_c)\times (1-r/r_c)$. Having tried other possible weight
functions, such as a parabolic shape $1-(r/r_C)^2$, or the quadratic shape
$(r/r_C)^2\times (1-r/r_C)$, we found that the thermalization and structure
properties are respectively close to the results obtained for the singular
$w(r)$ and regular $w_1(r)$ weight functions, emphasizing the importance of
the limit value $\lim_{r\to 0}w(r)/r$, as the key feature for reaping the best
of our second order scheme.

\section{Conclusion}

We demonstrated that a careful use of a Trotter splitting for Langevin
dynamics leads to a discretization scheme equivalent to the algorithm of
Ermak. The main difficulty in the procedure comes from the fact that the
``Liouvillian'' associated to the Langevin dynamics is time dependent, and the
contribution of the random noise in the stochastic equation, brings
significant changes compared with the case of an ordinary differential
equation.  As a consequence, it is necessary to take into account a commutator
term given by~(\ref{eq:TExpTripleCommutator}). For ordinary Langevin dynamics,
this corrective term simplifies to~(\ref{eq:decompositionEvolution}). This
term is responsible for the second independent Gaussian increment appearing in
the Ermak algorithm.

When checking how accurately these numerical schemes can reproduce a known
time-dependent correlation function, we found that keeping both independent
random Gaussian increments leads to better results for large integration time
steps. However, the benefits of the second random number gets smaller when the
integration time step is reduced, and in the small time step limit, where both
algorithms share the same accuracy, a single random Gaussian increment give
satisfactory results.

Using the same method, we were able to propose two algorithms for the
dissipative particle dynamics (DPD). One of them, termed ``ABA'', was found to
be easy to implement and efficient, and we recommend it as a reasonable
alternative for applications. Though less efficient, the other algorithm,
termed ``BAB'', deserves to be mentioned as an original and non trivial weak
order two algorithm. We observed that for simulations of soft core particles,
the unphysical structure of the pair correlation function $g(r)$ is
largely a consequence of a non vanishing weight function $w(r)$ near $r\to 0$.
This artefact almost disappears when the weight function vanishes at the
origin, without weakening the efficiency of the DPD thermostat. It is in this
limit only that high order algorithms give their best.

The theoretical developments exposed in this work are poised to provide
alternative derivations of some existing algorithms. Unlike the Fokker-Planck
way, our approach offers the possibility to discuss both strong and weak
convergence, a fact which extends its scope beyond the realm of MD
simulations.
 
The application to DPD revealed a situation of intermediate complexity,
standing between the usual additive noise Langevin dynamics, and the most
general multiplicative noise stochastic equations.  In particular, the
treatment of a Langevin dynamics with hydrodynamical interactions falls into
the same category as DPD, and our discussion will certainly help to improve
the design of numerical integration schemes, also in that case.


\appendix
\section{The sign convention}
\label{app:sign}

To convince ourselves of the importance of adopting the right convention 
when defining the Liouvillian, we consider a simple example.
We define the flow~1 as the ordinary differential equation (o.d.e):
\begin{equation}
\dot x = p;\;\; \dot p= 0;
\end{equation}
and the flow~2 as
\begin{equation}
\dot x = 0;\;\; \dot p= -p.
\end{equation}
Integrating these flows is easy and one finds:
\begin{equation} \Phi_1(x,p;t) = (x+pt, p); \end{equation}
and 
\begin{equation} \Phi_2(x,p;t) = (x, pe^{-t}). \end{equation}
Meanwhile, the Liouvillians associated to these flows are
\begin{equation} 
\Liou_1= p\frac{\partial}{\partial x}\;\;
\text{and}\;\;
\Liou_2= -p\frac{\partial}{\partial p}. 
\end{equation}
The commutator of $\Liou_1$ and $\Liou_2$ can be calculated, and its value is
\begin{equation} 
[\Liou_2,\Liou_1] = -p\frac{\partial}{\partial x}; 
\end{equation}
We now integrate along the flow~1 for a time interval of $\Delta t$ and along
the flow~2 for a time interval of $\Delta u$. We get the exact result
\begin{align}
\Phi_2(\Delta u) \circ\Phi_1(\Delta t)(x,p) &= 
  (x+p\Delta t, pe^{-\Delta u}),\\
\intertext{which, up to the order~2 in $\Delta t$, $\Delta u$ is equivalent
  to}  
\Phi_2(\Delta u)\circ\Phi_1(\Delta t)(x,p)&= \bigg(x+p\Delta t, p(1-\Delta u+\Delta
u^2/2)\bigg). 
\end{align}
Integrating the flows in the reverse order gives a slightly different result
\begin{align}
\Phi_1(\Delta t) \circ\Phi_2(\Delta u) (x,p) &= 
  (x+pe^{-\Delta u}\Delta t, pe^{-\Delta u}),\\
\intertext{which, up to the order~2 in $\Delta t$,$\Delta u$ is equivalent to} 
\Phi_1(\Delta t) \circ\Phi_2(\Delta u) (x,p) &= \bigg(x+p\Delta t-p\Delta
t\Delta u, p(1-\Delta u+\Delta u^2/2)\bigg). 
\end{align}
We observe that when commuting the flows, the following formula holds
\begin{equation}
\Phi_2(\Delta u) \circ\Phi_1(\Delta t)
=\Phi_{21}(\Delta u\Delta t)\circ\Phi_1(\Delta t) \circ\Phi_2(\Delta u),
\label{eq:commutatorFlow}
\end{equation}
with a flow $\Phi_{21}$ associated to the o.d.e
\begin{equation}
\dot x = p;\;\; \dot p = 0.
\end{equation}
and a Liouvillian $\Liou_{21}$ equal to
\begin{equation} \Liou_{21}= p\frac{\partial}{\partial x}; \end{equation}
In other words, $\Liou_{21}$ is the opposite of the usual commutator, for
operators acting on their right hand side:
\begin{equation}
\Liou_{21} = -[\Liou_2,\Liou_1].
\end{equation}
However, when dealing with the Liouvillians, the following formula holds:
\begin{equation}
e^{-\Delta u\Liou_2}e^{-\Delta t\Liou_1}\rho(x,p)
= e^{(-\Delta t)(-\Delta u)[\Liou_2,\Liou_1]}
e^{-\Delta t\Liou_1}e^{-\Delta u\Liou_2}\rho(x,p),
\label{eq:CommutatorLiouvillian}
\end{equation}
with, in particular, 
\begin{align}
e^{-\Delta t\Liou_1}\rho(x,p) &= \rho(x-p\Delta t,p)\notag\\
e^{-\Delta u\Liou_2}\rho(x,p) &= \rho(x,p e^{\Delta u})\notag\\
e^{-\Delta u\Liou_2}e^{-\Delta t\Liou_1}\rho(x,p)
&= \rho(x-p e^{\Delta u}\Delta t,p e^{\Delta u})\notag\\
e^{-\Delta t\Liou_1}e^{-\Delta u\Liou_2}\rho(x,p)
&= \rho(x-p\Delta t,p e^{\Delta u})\notag\\
e^{\Delta t\Delta u[\Liou_2,\Liou_1]}\rho(x-p\Delta t,p e^{\Delta u})
&= \rho(x-p\Delta t-p\Delta t\Delta u,p e^{\Delta u}).
\end{align}

We conclude that the algebraic formula~(\ref{eq:CommutatorLiouvillian})
involves the usual commutator of the differential operators (as, for instance,
in quantum mechanics), while the equivalent formula~(\ref{eq:commutatorFlow})
for the flows require a different definition of the commutator.

Thus, it is improper to write a flow as an exponential
\begin{equation}
\Phi(\Delta t)\equiv e^{\Delta t \Liou},
\label{eq:improperNotation}
\end{equation}
as it would imply 
\begin{equation}
e^{\Delta t\Liou_1}e^{\Delta u\Liou_2}=
e^{-\Delta u\Delta t[\Liou_1,\Liou_2]}
e^{\Delta u\Liou_2}e^{\Delta t\Liou_1},
\end{equation}
unless the operators appearing in the Liouvillian are meant to act on their
left hand side, such as $\frac{\overleftarrow{\partial}}{\partial x}$, 
$\frac{\overleftarrow{\partial}}{\partial p}$.
If, fortunately, in most cases the notation~(\ref{eq:improperNotation}) does
not have serious consequences, it cannot be used as soon as commutators are
explicitely required in the calculations.


\section{From chronological to regular exponentials: the Magnus expansion}
\label{app:Magnus}

We are now trying to evaluate 
\begin{equation}
\exp\left( \int_0^{\Delta t} \Liou(s)\dd s\right)
\cdot\mathrm{T}\exp\left( -\int_0^{\Delta t} \Liou(s)\dd s\right)
\label{eq:decompositionEvolution:2}
\end{equation}
We split $\Delta t$ in $n$ subintervals $\Delta t/n$ and write:
\begin{equation}
\mathrm{T}\exp\left( -\int_0^{\Delta t} \Liou(s)\dd s\right)
=\mathrm{T} \prod_{i=0}^{n-1} \exp\left( \int_{i\Delta t/n}^{(i+1)\Delta t/n}
  \Liou(s)\dd s\right) 
\end{equation}
Then, each chronological exponential is replaced by a usual exponential, and
the limit $n\to \infty$ of the product should converge towards a well defined
limit.
\begin{equation}
\mathrm{T}\exp\left( -\int_0^{\Delta t} \dd \Liou(s)\right)
=\lim_{n\to\infty} e^{-A_n}\cdot e^{-A_{n-1}}\cdot\ldots\cdot
e^{-A_2}\cdot e^{-A_1}
\end{equation}
where $A_i$ is simply 
\begin{equation}
e^{-A_i}=\exp\left(-\int_{(i-1)\Delta t/n}^{i\Delta t/n} \dd s\,\Liou(s)
\right) 
\end{equation}
Thus, the product~(\ref{eq:decompositionEvolution:2}) can be rewritten as
\begin{equation}
\lim_{n\to\infty} 
e^{A_1+A_2+\ldots+A_{n-1}+A_{n}}\cdot
e^{-A_n}\cdot e^{-A_{n-1}}\cdot\ldots\cdot e^{-A_2}\cdot e^{-A_1}
\label{eq:manyASplitting}
\end{equation}
The last exponential can be further reduced by using the BCH formula,
keeping the triple commutators:
\begin{eqnarray}
\exp(A)\exp(B)&=&\exp\left(A+B+\frac{1}{2}[A,B]+\frac{1}{12}[A,[A,B]]
+\frac{1}{12}[B,[B,A]]\ldots\right)
\label{eq:BCHDirect3:3}
\end{eqnarray}
Using BCH again on the right hand side of~(\ref{eq:BCHDirect3:3}), we find a
reverse formula, exact up to triple commutators,
\begin{equation}
\exp(A+B)=\exp\left(\frac{1}{2}[B,A]-\frac{1}{3}[A,[A,B]]
+\frac{1}{6}[B,[B,A]]\right)\exp(A)\exp(B)
\label{eq:BCHReverse3:3}
\end{equation}
To reduce the last term in~(\ref{eq:manyASplitting}), we define the sums 
\begin{eqnarray}
S_1 &=& 0;\nonumber\\
S_i &=& \sum_{j=1}^{i-1} A_j =A_{1}+A_{2}+\ldots+A_{i-1},
\end{eqnarray}
and we apply~(\ref{eq:BCHReverse3:3}) recursively to 
\begin{eqnarray}
e^{A_1+A_2+\ldots+A_n} &=& e^{S_n+A_n}\nonumber\\
&=& e^{-\frac{1}{2}[S_n,A_n]-\frac{1}{3}[S_n,[S_n,A_n]]
  +\frac{1}{6}[A_n,[A_n,S_n]]}\cdot 
  e^{S_{n}=A_{n-1}+S_{n-1}}\cdot e^{A_n}
\nonumber\\
&=& e^{-\frac{1}{2}[S_n,A_n]-\frac{1}{3}[S_n,[S_n,A_n]]
  +\frac{1}{6}[A_n,[A_n,S_n]]}\nonumber\\
& & 
\cdot e^{ -\frac{1}{2}[S_{n-1},A_{n-1}]-\frac{1}{3}[S_{n-2},[S_{n-2},A_{n-2}]] 
 +\frac{1}{6}[A_{n-2},[A_{n-2},S_{n-2}]]}
 \cdot e^{S_{n-1}}\cdot e^{A_{n-1}}\cdot e^{A_{n}} \nonumber\\
&=&  e^{-\frac{1}{2}\sum_{i=1}^{n}[S_i,A_i]
  -\frac{1}{3}\sum_{i=1}^{n}[S_i,[S_i,A_i]] 
  +\frac{1}{6}\sum_{i=1}^{n}[A_i,[A_i,S_i]]+R_n}
   \cdot e^{A_1}\cdot e^{A_2}\cdot\ldots e^{A_{n}}.
\end{eqnarray}

In the large $n$ limit, $A_i\sim \sqrt{\Delta t}/n$, $S_i\sim \sqrt{\Delta t}$
and $R_n\sim 1/n$, it results that the term
$\frac{1}{6}\sum_{i=1}^{n}[A_i,[A_i,S_i]]$ and the remaining term $R_n$ do not
contribute.  Equation~(\ref{eq:manyASplitting}) reduces to
\begin{equation}
\lim_{n\to\infty} \exp\left(-\frac{1}{2}\sum_{i=1}^{n}[S_i,A_i]
 -\frac{1}{3}\sum_{i=1}^{n}[S_i,[S_i,A_i]]\right),
\end{equation}
or, using Riemann sums instead,
\begin{equation}
\exp\left(-\frac{1}{2}\int_{0}^{\Delta t} \dd s\int^{\displaystyle s}_{0} 
 \dd u [\Liou(u),\Liou(s)] -\frac{1}{3} \int_{0}^{\Delta t} \dd s
 \int_0^{\displaystyle s} \dd u \int_0^{\displaystyle s} \dd v
 [\Liou(v),[\Liou(u),\Liou(s)]]\right).
\label{eq:TExpTripleCommutatorB}
\end{equation}
When specifying to the Langevin dynamics, 
\begin{equation}
\dd s\, \Liou(s) = \dd s \left( \frac{p}{m}\frac{\partial}{\partial r}
+(f(r)-\gamma p)\frac{\partial}{\partial p} +\sigma \psi(s)
\frac{\partial}{\partial p}\right),
\end{equation}
we obtain 
\begin{equation}
[\dd s\, \Liou(s),\dd u\, \Liou(u)] =
(\psi(u)\dd u\dd s\, -\psi(s)\dd s\dd u)
\left( \frac{1}{m}\frac{\partial}{\partial r}
-\gamma \frac{\partial}{\partial p}\right),
\end{equation}
which contributes to a $\Delta t^{3/2}$ term, while the triple commutator
gives only a $\Delta t^{5/2}$ term and can be disregarded. Finally,
\begin{eqnarray}
-\frac{1}{2}\int_{0}^{\Delta t} \dd s \int_0^{\displaystyle s} 
  \dd u  [\Liou(u),\Liou(s)]
&=& \frac{1}{2} \int_0^{\Delta t}\left(s\,\psi(s)\dd s\,-\dd
  s\,\Wien{0}{s})\right)\left( \frac{1}{m}\frac{\partial}{\partial r}
-\gamma \frac{\partial}{\partial p}\right);\nonumber\\
&=& \bigg[\frac{\Delta t}{2} \Wien{0}{\Delta t} -\int_0^{\Delta t}\dd s\,
  \Wien{0}{s}\bigg]\notag\\
& & 
\times \left( \frac{1}{m}\frac{\partial}{\partial r}
-\gamma \frac{\partial}{\partial p}\right),
\end{eqnarray}
and the justification of (\ref{eq:decompositionEvolution}) is complete.

The Magnus expansion cited in eq.~(14) of Ref.~\cite{BurBur99} is a commutator
expansion for the logarithm of a chronological exponential:
\begin{equation}
\Omega(t) = \ln \left(\mathrm{T}
  \exp\left(\int_{0}^{t}\dd s\,\Liou(s)\right)\right),
\end{equation}
The three first terms of the expansion are 
\begin{eqnarray}
\Omega(t) &=& \int_{0}^{t}\dd s\,\Liou(s) 
+\frac{1}{2}\int_{0}^{t}\dd s_1\,\int_{0}^{s_1}\dd s_2\,
  [\Liou(s_1),\Liou(s_2)] \notag\\
& & +\frac{1}{4}\int_{0}^{t}\dd s_1\,
  \int_{0}^{s_1}\dd s_2\,  \int_{0}^{s_2}\dd s_3\,
  [\Liou(s_1),[\Liou(s_2),\Liou(s_3)]] \notag\\
& & +\frac{1}{12}\int_{0}^{t}\dd s_1\,
 \int_{0}^{s_1}\dd s_2\,\int_{0}^{s_1}\dd s_3\,
  [[\Liou(s_1),\Liou(s_2)],\Liou(s_3)].\label{eq:realMagnusExpansion}
\end{eqnarray}
Using BCH, we get
\begin{eqnarray}
\exp(\Omega(t)) &=& \exp\left(\int_{0}^{t}\dd s\,\Liou(s) \right)\cdot
\exp \left(\frac{1}{2}\int_{0}^{t}\dd s_1\,\int_{0}^{s_1}\dd s_2\,
  [\Liou(s_1),\Liou(s_2)]\right) \notag\\
& & \cdot\exp\bigg( +\frac{1}{4}\int_{0}^{t}\dd s_1\,
  \int_{0}^{s_1}\dd s_2\,  \int_{0}^{s_2}\dd s_3\,
  [\Liou(s_1),[\Liou(s_2),\Liou(s_3)]] \notag\\
& & +\frac{1}{12}\int_{0}^{t}\dd s_1\,
 \int_{0}^{s_1}\dd s_2\,\int_{0}^{s_1}\dd s_3\,
  [[\Liou(s_1),\Liou(s_2)],\Liou(s_3)]\notag\\
& & +\frac{1}{2}\left[\frac{1}{2}\int_{0}^{t}\dd
    s_1\,\int_{0}^{s_1}\dd s_2\, [\Liou(s_1),\Liou(s_2)],\int_{0}^{t}\dd
    s_3\,\Liou(s_3)\right] \bigg).
\label{eq:realMagnusExpansion:2}
\end{eqnarray}
Finally, the last commutator reads 
\begin{multline}
\frac{1}{4}\int_{0}^{t}\dd s_1\,
  \int_{0}^{s_1}\dd s_2\,  \int_{0}^{t}\dd s_3\,
  [[\Liou(s_1),\Liou(s_2)],\Liou(s_3)]=\\
\frac{1}{4}\int_{0}^{t}\dd s_1\,
  \int_{0}^{s_1}\dd s_2\,\int_{0}^{s_1}\dd s_3\,
 [[\Liou(s_1),\Liou(s_2)],\Liou(s_3)]\\
 +\frac{1}{4}\int_{0}^{t}\dd s_1\,
  \int_{0}^{s_1}\dd s_2\,\int_{s_1}^{t}\dd s_3\,
 [[\Liou(s_1),\Liou(s_2)],\Liou(s_3)]\\
= \frac{1}{4}\int_{0}^{t}\dd s_1\,
  \int_{0}^{s_1}\dd s_2\,\int_{0}^{s_1}\dd s_3\,
 [[\Liou(s_1),\Liou(s_2)],\Liou(s_3)]\\
 -\frac{1}{4}\int_{0}^{t}\dd s_3\,
  \int_{0}^{s_3}\dd s_1\,\int_{s_1}^{t}\dd s_2\,
 [\Liou(s_3),[\Liou(s_1),\Liou(s_2)]].
\label{eq:realMagnusExpansion:3}
\end{multline}
The last term in~(\ref{eq:realMagnusExpansion:3}) cancels the second line
of~(\ref{eq:realMagnusExpansion}), and one gets, after renaming the dummy
integration variables:
\begin{eqnarray}
\exp(\Omega(t)) &=& \exp\left(\int_{0}^{t}\dd s\,\Liou(s) \right)\cdot
\exp \bigg(\frac{1}{2}\int_{0}^{t}\dd s \,\int_{0}^{s }\dd u\,
  [\Liou(s),\Liou(u)] \notag\\
& & +\frac{1}{3}\int_{0}^{t}\dd s\,
 \int_{0}^{s}\dd u\,\int_{0}^{s}\dd v\,
  [[\Liou(s),\Liou(u)],\Liou(v)]\bigg),
\label{eq:realMagnusExpansion:4}
\end{eqnarray}
or equivalently,
\begin{eqnarray}
\exp(-\Omega(t)) &=& \exp\left(-\int_{0}^{t}\dd s\,\Liou(s) \right)\cdot
\exp \bigg(-\frac{1}{2}\int_{0}^{t}\dd s \,\int_{0}^{s }\dd u\,
  [\Liou(u),\Liou(s)] \notag\\
& & -\frac{1}{3}\int_{0}^{t}\dd s\,
 \int_{0}^{s}\dd u\,\int_{0}^{s}\dd v\,
  [\Liou(v),[\Liou(u),\Liou(s)]]\bigg),
\label{eq:realMagnusExpansion:5}
\end{eqnarray}
identical to~(\ref{eq:TExpTripleCommutatorB}).

\section{The Velocity-Verlet and Langevin algorithms}
\label{app:vv}

Let us start with the splitting~(\ref{eq:splittingVV:1}). 
Each operator acts recursively on the coordinates $r,p$
of $\rho(r,p)$. This leads to the sequence:
\begin{eqnarray}
\exp(-A/2)\,\rho(r,p) &=& \rho\left(r-p\frac{\Delta t}{2m},p\right)
\nonumber\\
\exp(-B)\,\rho(r,p) &=& \rho\left(r,p-f(r)\Delta t \right)
\nonumber\\
\exp(-B)\,\exp(-A/2)\,\rho(r,p) &=&
\rho\left(r-(p-f(r)\Delta t)\frac{\Delta t}{2m}, 
  p-f(r)\Delta t\right) \nonumber\\ 
\exp(-A/2)\exp(-B)\exp(-A/2)\,\rho(r,p)&=& \nonumber\\
 & & \hspace{-4cm}
\rho\left( r-\frac{\Delta t}{2m}\bigg[p + p -\Delta t 
f\bigg(r-p\frac{\Delta t}{2m}\bigg)\bigg],
 p-\Delta t f\bigg(r-p\frac{\Delta t}{2m}\bigg)
 \right).
\end{eqnarray}
The arguments of $\rho$ must be identified to the initial position and
impulsion $(r(t),p(t))$ of a trajectory ending at the final position
$r=r(t+\Delta t)$, $p=p(t+\Delta t)$. Thus,
\begin{eqnarray}
p(t) &=& p(t+\Delta t)-f\left(r(t+\Delta t)-p(t+\Delta t)\frac{\Delta
  t}{2m}\right)\Delta t; \label{eq:val:1}\\ 
r(t) &=& r(t+\Delta t) -(p(t+\Delta t)+p(t))\frac{\Delta t}{2m}.
\end{eqnarray}
Then, equations can be reversed to express $(r(t+\Delta t), p(t+\Delta t))$ as
functions of $(r(t),p(t))$, and in~(\ref{eq:val:1}), 
$r(t+\Delta t)-p(t+\Delta t)\Delta t/2m$ can be  replaced by 
$r(t)+p(t)\Delta t/2m$. 
\begin{eqnarray}
p(t+\Delta t) &=& p(t)+f\left(r(t)+p(t)\frac{\Delta t}{2m}\right)
\Delta t;\\ 
r(t+\Delta t) &=& r(t)+(p(t)+p(t+\Delta t))\frac{\Delta t}{2m}.
\end{eqnarray}
Meanwhile, the splitting~(\ref{eq:splittingVV:2}) leads to
\begin{eqnarray}
p(t+\Delta t) &=& p(t)+\frac{\Delta t}{2}
\bigg(f(r(t))+f(r(t+\Delta t))\bigg);\\ 
r(t+\Delta t) &=& r(t)+ p(t)\frac{\Delta t}{m} + f(r(t))\frac{\Delta t^2}{2m}.
\end{eqnarray}
For the Langevin dynamics of one dimensional systems, we use:
\begin{multline}
\exp(-A/2)\exp(-B)\exp(-A/2)\cdot
\exp\bigg\lbrace\sigma\bigg\lbrack
 \frac{\Delta t}{2}\Wien{t}{t+\Delta t}\\
 -\int_t^{t+\Delta t} \dd s\,\Wien{t}{s}\bigg\rbrack 
\left(\frac{1}{m}\frac{\partial}{\partial r} 
  -\gamma\frac{\partial}{\partial p}\right)
\bigg\rbrace \rho(r,p),
\end{multline}
S%
with $A=\Delta t\,p \frac{\partial}{\partial r}$; $B=[\Delta t\, (f(r) -\gamma
p) +\sigma\Wien{t}{t+\Delta t}] \frac{\partial}{\partial p}$.  
Given that
\begin{equation}
\exp\bigg[(X(r) + Y p)\frac{\partial}{\partial p}\bigg]\rho(r,p)
 = \rho\left(r,p e^Y + X(r)\frac{e^Y-1}{Y} \right),
\end{equation}
we find
\begin{eqnarray}
e^{-A/2}\rho(r,p) &=& \rho\left(r-p\frac{\Delta t}{2m},p\right)\nonumber\\
e^{-B}\rho(r,p) &=& \rho\left(r,p-\left(\Delta t f(r)+\sigma\Wien{t}{t+\Delta
      t} \right)\frac{e^{\gamma\Delta t}-1}{\gamma \Delta t}\right)
\nonumber\\
e^{-A/2}e^{-B}e^{-A/2} \rho(r,p) &=& 
\rho\left( r -\frac{\Delta t}{2m}\bigg[ p+ pe^{\gamma \Delta t}
-\bigg(\Delta t f(r) +\sigma\Wien{t}{t+\Delta t}\bigg)
\frac{e^{\gamma \Delta t}-1}{\gamma \Delta t}\bigg],\right.
\nonumber\\
& &  \left. pe^{\gamma \Delta t}
-\bigg(\Delta t f(\tilde{r}) +\sigma\Wien{t}{t+\Delta t}\bigg)
\frac{e^{\gamma \Delta t}-1}{\gamma \Delta t}\right).
\end{eqnarray}
In the above expression, $r$ must be identified with $r(t+\Delta t)$, and $p$
with $p(t+\Delta t)$. The resulting function is
$e^{-A/2}e^{-B}e^{-A/2}\rho(r,p) = \rho(r',p')$, where $r'$ must be replaced
with $r(t)$, $p'$ with $p(t)$, and $r'$, $p'$ given by the above expression,
with $\tilde{r}\simeq r(t)+p(t)/2m$ at the desired level of accuracy. These
relations can be reverted in such a way that $p(t+\Delta t)$ and $r(t+\Delta
t)$ depend explicitely on $p(t)$ and $r(t)$. The result is
\begin{eqnarray}
p(t+\Delta t) &=& p(t)e^{-\gamma \Delta t}
 +\bigg[ \Delta t f\bigg(r(t)+p(t)\frac{\Delta t}{2m}\bigg) 
 +\sigma\Wien{t}{t+\Delta t}\bigg]\frac{1-e^{-\gamma\Delta t}}{\gamma\Delta t} 
\nonumber\\
r(t+\Delta t) &=& r(t)+\frac{\Delta t}{2m}\bigg(p(t)+p(t+\Delta t)\bigg)
\end{eqnarray}
Finally, the commutator correction simply shifts the previous result by a
$\Delta t^{3/2}$ amount. 
\begin{eqnarray}
p(t+\Delta t) &=& p(t)e^{-\gamma\Delta t} 
+\bigg[\Delta t f\bigg(r(t)+p(t)\frac{\Delta t}{2m}\bigg)
+\sigma\Wien{t}{t+\Delta t}\bigg]
\frac{1-e^{-\gamma\Delta t}}{\gamma\Delta t}\nonumber\\
& & +\gamma\sigma\bigg[\frac{\Delta t}{2}\Wien{t}{t+\Delta t}
 -\int_0^{\Delta t}\dd s\,\Wien{t}{s}\bigg];\\
r(t+\Delta t) &=& r(t)+\bigg(p(t)+p(t+\Delta t)\bigg)
  \frac{\Delta t}{2m}\nonumber\\
& & +\frac{\sigma}{m}\bigg[\int_0^{\Delta t}\dd s\,\Wien{t}{s}
  -\frac{\Delta t}{2}\Wien{t}{t+\Delta t}\bigg]. 
\end{eqnarray}
Alternatively, the other splitting leads to slightly more complex 
expressions:
\begin{eqnarray}
e^{-B/2}e^{-A}e^{-B/2} \rho(r,p) &=& 
\rho\left( r -\frac{\Delta t}{m}\bigg[ pe^{\gamma \Delta t/2}
-\bigg(\Delta t f(r)+\sigma\Wien{t}{t+\Delta t}\bigg) 
\frac{e^{\gamma \Delta t/2}-1}{\gamma\Delta t}\bigg]
,\right.
\nonumber\\
& &   pe^{\gamma \Delta t} 
-\bigg(\Delta t f(r) +\sigma\Wien{t}{t+\Delta t}\bigg)
 e^{\gamma \Delta t/2} \frac{e^{\gamma \Delta t/2}-1}{\gamma\Delta t}
\nonumber\\
& & \left. -\bigg(\Delta t f(\tilde{r}) +\sigma\Wien{t}{t+\Delta t}\bigg)
\frac{e^{\gamma \Delta t/2}-1}{\gamma\Delta t}
\right)
\end{eqnarray}
This finally leads to the algorithm
\begin{eqnarray}
p(t+\Delta t) &=& p(t)e^{-\gamma\Delta t}+\gamma^{-1}(1-e^{-\gamma \Delta t/2})
\bigg(f(r(t+\Delta t))+f(r(t))e^{-\gamma\Delta t/2}\bigg)
\nonumber\\
& & 
+\sigma\Wien{t}{t+\Delta t}\frac{1-e^{-\gamma\Delta t}}{\gamma\Delta t}
\nonumber\\
& & +\gamma\sigma\bigg[\frac{\Delta t}{2}\Wien{t}{t+\Delta t}
 -\int_0^{\Delta t}\dd s\,\Wien{t}{s}\bigg];\\
r(t+\Delta t) &=& r(t)+\frac{\Delta t}{m}
\bigg[p(t)e^{-\gamma \Delta t/2}+\frac{1-e^{-\gamma\Delta t/2}}{\gamma\Delta t}
\bigg(\Delta t f(r(t))+\sigma\Wien{t}{t+\Delta t}
\bigg)\bigg]
\nonumber\\
& & +\frac{\sigma}{m}\bigg[\int_0^{\Delta t}\dd s\,\Wien{t}{s}
  -\frac{\Delta t}{2}\Wien{t}{t+\Delta t}\bigg]. 
\end{eqnarray}
These results generalize straightforwardly to the three dimensional many
particles case, by adding the space and particles indices. For instance
\begin{eqnarray}
\bp_{i\alpha}(t+\Delta t) &=& \bp_{i\alpha}(t)e^{-\gamma\Delta t} 
+\bigg[\Delta t \bf_{i\alpha}\bigg(\br(t)+\bp(t)
\frac{\Delta t}{2m}\bigg) +\sigma\Wieni{t}{t+\Delta t}{i\alpha}\bigg]
\frac{1-e^{-\gamma\Delta t}}{\gamma\Delta t}\nonumber\\
& & +\gamma\sigma\bigg[\frac{\Delta t}{2}\Wieni{t}{t+\Delta t}{i\alpha}
 -\int_0^{\Delta t}\dd s\,\Wieni{t}{s}{i\alpha}\bigg];\\
\br_{i\alpha}(t+\Delta t) &=&
  \br_{i\alpha}(t)+\bigg(\bp_{i\alpha}(t)+\bp_{i\alpha}(t+\Delta t)\bigg) 
  \frac{\Delta t}{2m}\nonumber\\
& & +\frac{\sigma}{m}\bigg[\int_0^{\Delta t}\dd s\,\Wieni{t}{s}{i\alpha}
  -\frac{\Delta t}{2}\Wieni{t}{t+\Delta t}{i\alpha}\bigg]. 
\end{eqnarray}
corresponds to the splitting~(\ref{eq:splitting13}). 
If one expands in powers of $\Delta t$ up to $\Delta t^2$, the algorithm reads
\begin{eqnarray}
p_{i\alpha}(t+\Delta t) &=& p_{i\alpha}(t)
 \left(1-\gamma\Delta t +\frac{\gamma^2\Delta  t^2}{2}\right) 
  +f_{i\alpha}\left(\br(t)+\bp(t)\frac{\Delta t}{2m}\right)
  \Delta t\left(1-\frac{\gamma\Delta t}{2}\right)\nonumber\\
& & +\sigma\Wieni{t}{t+\Delta t}{i\alpha}
  -\sigma\gamma\int_t^{t+\Delta t}\dd s\,\Wieni{t}{s}{i\alpha};
  \label{eq:splittingABAP:2}\\  
r_{i\alpha}(t+\Delta t) &=& r_{i\alpha}(t)+p_{i\alpha}(t)
 \frac{\Delta t}{m}\left(1-\frac{\gamma \Delta t}{2}\right) +
  f_{i\alpha}(\br(t))\frac{\Delta t^2}{2m}\nonumber\\
& &  +\frac{\sigma}{m}\int_t^{t+\Delta t}\dd s\,\Wieni{t}{s}{i\alpha},
\label{eq:splittingABAR:2}
\end{eqnarray}
which are eq~(\ref{eq:splittingABAR}) and~(\ref{eq:splittingABAP}).
Alternatively, the splitting~(\ref{eq:splitting14}) leads to:
\begin{eqnarray}
\bp_{i\alpha}(t+\Delta t) &=& \bp_{i\alpha}(t)e^{-\gamma\Delta t}
+\gamma^{-1}(1-e^{-\gamma \Delta t/2})
\bigg(\bf_{i\alpha}(\br(t+\Delta t))
 +\bf_{i\alpha}(\br(t))e^{-\gamma\Delta t/2}\bigg) 
\nonumber\\
& & 
+\sigma\Wieni{t}{t+\Delta t}{i\alpha} 
\frac{1-e^{-\gamma\Delta t}}{\gamma\Delta t}
\nonumber\\
& & +\gamma\sigma\bigg[\frac{\Delta t}{2}\Wieni{t}{t+\Delta t}{i\alpha}
 -\int_0^{\Delta t}\dd s\,\Wieni{t}{s}{i\alpha}\bigg];\\
\br_{i\alpha}(t+\Delta t) &=& \br_{i\alpha}(t)+\frac{\Delta t}{m}
\bigg[\bp_{i\alpha}(t)e^{-\gamma \Delta t/2}
+\frac{1-e^{-\gamma\Delta t/2}}{\gamma\Delta t}
\bigg(\Delta t \bf_{i\alpha}(\br(t))+\sigma\Wieni{t}{t+\Delta t}{i\alpha}
\bigg)\bigg]
\nonumber\\
& & +\frac{\sigma}{m}\bigg[\int_0^{\Delta t}\dd s\,\Wieni{t}{s}{i\alpha}
  -\frac{\Delta t}{2}\Wieni{t}{t+\Delta t}{i\alpha}\bigg]. 
\end{eqnarray}
and finally,
\begin{eqnarray}
\bp_{i\alpha}(t+\Delta t) &=& \bp_{i\alpha}(t)
 \bigg(1-\gamma\Delta t +\frac{\gamma^2\Delta t^2}{2}\bigg) 
 +\frac{\Delta t}{2}(\bf_{i\alpha}(\br(t))+\bf_{i\alpha}(\br(t+\Delta t)) 
 -\gamma \frac{\Delta t^2}{2} \bf_{i\alpha}(\br(t))
\nonumber\\
& & +\sigma\Wieni{t}{t+\Delta t}{i\alpha}
 -\gamma\sigma\int_t^{t+\Delta t}\dd s\,\Wieni{t}{s}{i\alpha};
\label{eq:splittingBABP:2}\\ 
\br_{i\alpha}(t+\Delta t) &=& \br_{i\alpha}(t) + \bp_{i\alpha}
\frac{\Delta t}{m}\bigg(1-\gamma\frac{\Delta t}{2}\bigg)  
 +\bf_{i\alpha}(\br(t))\frac{\Delta t^2}{2m} 
 +\frac{\sigma}{m}\int_t^{t+\Delta t}\!\dd s\,\Wieni{t}{s}{i\alpha},
\label{eq:splittingBABR:2} 
\end{eqnarray}
which are eq~(\ref{eq:splittingBABP}) and~(\ref{eq:splittingBABR}).


\section{Derivation of the DPD algorithm}
\label{app:dpd}

The operators $\exp(-A)$ and $\exp(-B)$, with $A$ and $B$ introduced
in~(\ref{eq:definitionAB:dpd}), act respectively on test functions like
\begin{eqnarray}
e^{-B} \rho(\br,\bp) &=& \rho(\br, e^{\Delta t \bGamma(\br)}\cdot\bp - 
  \bGamma(\br)^{-1}(e^{\Delta t\,\bGamma(\br)}-\mathbf{1})\cdot
  \bK(\br;t)\,)\\ 
e^{-A} \rho(\br,\bp) &=& \rho(\br-\frac{\Delta t}{m}\bp,\bp).
\end{eqnarray}
The splitting $e^{-A/2}e^{-B}e^{-A/2}$ leads to, following manipulations
similar to appendix~(\ref{app:vv}):
\begin{eqnarray}
\br(t+\Delta t) &=& \br(t)+\bigg(\bp(t)+\bp(t+\Delta t)\bigg)\frac{\Delta
  t}{2m}; \\
\bp(t+\Delta t) &=& 
e^{-\Delta t\bGamma(\br(t)+\bp(t)\frac{\Delta t}{2m})}\bp(t)\notag\\
& & \hspace{-1cm}
+\bGamma\bigg(\br(t)+\bp(t)\frac{\Delta t}{2m}\bigg)^{-1}
\bigg(\bm{1}-e^{-\Delta t\bGamma(\br(t)+\bp(t)\frac{\Delta t}{2m})}\bigg)
\cdot\bK\bigg(\br(t)+\bp(t)\frac{\Delta t}{2m};t\bigg),
\label{eq:exampleReversible}
\end{eqnarray}
which, upon expansion in powers of $\Delta t$ up to $\Delta t^2$, gives the
algorithm~(\ref{eq:splittingDPDABA:P}) and~(\ref{eq:splittingDPDABA:R}).
The second splitting $e^{-B/2}e^{-A}e^{-B/2}$ leads to:
\begin{eqnarray}
\br(t+\Delta t) &=& \br(t)+\frac{\Delta t}{m}
\bigg[ e^{-\frac{\Delta t}{2}\bGamma(\br(t))}\bp(t)
+\bGamma(\br(t))^{-1}\bigg(\bm{1}-e^{-\frac{\Delta t}{2}\bGamma(\br(t))}\bigg)
\cdot\bK(r(t);t)\bigg]; \\
\bp(t+\Delta t) &=& e^{-\frac{\Delta t}{2}\bGamma(\br(t+\Delta t))}
e^{-\frac{\Delta t}{2}\bGamma(\br(t))} \bp(t)\notag\\
& & +e^{-\frac{\Delta t}{2}\bGamma(\br(t+\Delta t))}
\bGamma(\br(t))^{-1}\bigg(\bm{1}-e^{-\frac{\Delta t}{2}\bGamma(\br(t))}\bigg) 
\cdot\bK(r(t);t)\notag\\
& & +\bGamma(\br(t+\Delta t))^{-1}
\bigg(\bm{1}-e^{-\frac{\Delta t}{2}\bGamma(\br(t+\Delta t))}\bigg) 
\cdot\bK(r(t+\Delta t);t),
\end{eqnarray}
which, upon expansion in powers of $\Delta t$ up to $\Delta t^2$, gives the
algorithm~(\ref{eq:splittingDPDBAB:P}) and~(\ref{eq:splittingDPDBAB:R}).
The correction term $[B,[A,B]]$ acts only on the impulsions:
\begin{eqnarray}
[B,[A,B]] &=& 2\frac{\sigma^2}{m}\Delta t
\sum_{ijkl,\alpha\beta} 
\dpdWien{t}{t+\Delta t}{ij}\dpdWien{t}{t+\Delta t}{kl}\notag\\
& & \times
\frac{w(r_{ij})}{r_{ij}}(r_{i\alpha}-r_{j\alpha}) 
\frac{\partial}{\partial r_{i\alpha}} \Big(  
\frac{w(r_{kl})}{r_{kl}}(r_{k\beta}-r_{l\beta})\Big)
\frac{\partial}{\partial p_{k\beta}}, 
\label{eq:correctionDeltaP1}
\end{eqnarray}
As a further simplification, we remark that the terms involving two different
pairs $ij$ and $kl$ are of zero mean, uncorrelated from any
$\dpdWien{t}{t+\Delta t}{mn}$, and of variance $\Delta t^4$. Therefore, within
a weak order two scheme, we can replace~(\ref{eq:correctionDeltaP1}) by its
average value and the correction term becomes:
\begin{equation}
[B,[A,B]]= 2\frac{\sigma^2}{m}\Delta t^2 
\sum_{ij,\alpha\beta}
\frac{w(r_{ij})}{r_{ij}}(r_{i\alpha}-r_{j\alpha})
\frac{\partial}{\partial r_{i\alpha}} 
\Big(\frac{w(r_{ij})}{r_{ij}}(r_{i\beta}-r_{j\beta})\Big)
\left(\frac{\partial}{\partial p_{i\beta}}
 -\frac{\partial}{\partial p_{j\beta}}\right),
\label{eq:DPDBAB}
\end{equation}
and finally, each impulsion $p_{i\beta}$ is incremented by an amount 
\begin{equation}
(\Delta p_{i\beta})_1=\frac{4\sigma^2\Delta t^2}{mc}
\sum_{j\neq i,\alpha}\frac{w(r_{ij})}{r_{ij}}(r_{i\alpha}-r_{j\alpha}) 
\frac{\partial}{\partial r_{i\alpha}} 
\Big(\frac{w(r_{ij})}{r_{ij}}(r_{i\beta}-r_{j\beta})\Big),
\label{eq:incrementDPDBAB}
\end{equation}
where $c$ is 12 for the splitting $e^{-A/2}e^{-B}e^{-A/2}$ or -24
for the splitting $e^{-B/2}e^{-A}e^{-B/2}$.

The correction $\exp(-1/2\iint\dd u\,\dd s\,[\Liou(u),\Liou(s)])$ gives
terms proportional to \hbox{$\int (s-\Delta t/2)\psi_{ij}(s)\dd s$}, which
have been shown to be uncorrelated to the $\dpdWien{t}{t+\Delta t}{ij}$, of
vanishing mean value and of variance $\Delta t^3$. These terms are the pendant
of the corrections in the Ermak algorithm, and can be neglected in practice.

The correction term $\exp(-1/3\iiint\dd v\,\dd u\,\dd
s\,[\Liou(v),[\Liou(u),\Liou(s)]])$ can be written as
\begin{multline}
\displaystyle
-\frac{\sigma^2}{3m}\sum_{i,j,k,l}\int_t^{t+\Delta t} \Big(\dpdWien{t}{s}{ij}
 \psi_{kl}(s)(s-t)\dd s\,+\dpdWien{t}{s}{kl}\psi_{ij}(s)(s-t)\dd s\notag\\
-2\dd s \dpdWien{t}{s}{ij} \dpdWien{0}{s}{kl}\Big)\\
\displaystyle\sum_{\alpha,\beta}\left\lbrace
\frac{w(r_{ij})}{r_{ij}}(r_{i\alpha}-r_{j\alpha})
\frac{\partial}{\partial r_{i\alpha}} \Big(\frac{w(r_{kl})}{r_{kl}}
(r_{k\beta}-r_{l\beta})\Big)\frac{\partial}{\partial p_{k\beta}}
\right\rbrace. 
\label{eq:correctionDPDtriple}
\end{multline}
If the pairs $(i,j)$ and $(k,l)$ differ, the random increment turns out to be
uncorrelated to the $\dpdWien{t}{t+\Delta t}{ij}$, of variance $\Delta t^4$ and
therefore can be evacuated from an order two integration scheme. However, the
term corresponding to the same pairs gives an positive $\Delta t^2$
contribution (we use here the Stratonovitch interpretation of the stochastic
integrals). 
\begin{multline}
\displaystyle\mean{\int_t^{t+\Delta t} \Big(\dpdWien{t}{s}{ij}
 \psi_{ij}(s)(s-t)\dd s\, -\dd s\dpdWien{t}{s}{ij}^2\Big)}\\
= \displaystyle\mean{
\frac{\Delta t}{2} \dpdWien{t}{t+\Delta t}{ij}^2
 -\frac{3}{2}\int_t^{t+\Delta t}\dd s 
\dpdWien{t}{s}{ij}^2} =\frac{-\Delta t^2}{4},
\end{multline}
while the last term of equation~(\ref{eq:correctionDPDtriple}) reduces to
(\ref{eq:DPDBAB}), leading to an operator:
\begin{multline}
\displaystyle
-\frac{2\sigma^2}{3m}\sum_{i,j}\mean{\int_t^{t+\Delta t}
 \bigg(\dpdWien{t}{s}{ij}
 \psi_{ij}(s)(s-t)\dd s\,-\dd s \dpdWien{0}{s}{ij}^2\bigg)}\\
\times\displaystyle\sum_{\alpha,\beta}\left\lbrace
\frac{w(r_{ij})}{r_{ij}}(r_{i\alpha}-r_{j\alpha})
\frac{\partial}{\partial r_{i\alpha}} \Big(\frac{w(r_{ij})}{r_{ij}}
(r_{i\beta}-r_{j\beta})\Big)
\bigg(\frac{\partial}{\partial p_{i\beta}}-\frac{\partial}{\partial
  p_{j\beta}}\bigg)
\right\rbrace, 
\end{multline}
and the corresponding corrective term for the impulsions
$p_{i\beta}$ equals
\begin{equation}
(\Delta p_{i\beta})_2=-\frac{\sigma^2\Delta t^2}{3m}
\sum_{j,(j\neq i),\alpha}\frac{w(r_{ij})}{r_{ij}}(r_{i\alpha}-r_{j\alpha}) 
\frac{\partial}{\partial r_{i\alpha}} 
\Big(\frac{w(r_{ij})}{r_{ij}}(r_{i\beta}-r_{j\beta})\Big).
\label{eq:incrementDPDTriple}
\end{equation}
Together, the correction of the triple commutators is the sum of
$(\Delta p_{i\beta})_1$~(\ref{eq:incrementDPDBAB}) and
$(\Delta p_{i\beta})_2$~(\ref{eq:incrementDPDTriple}) are
\begin{equation}
\frac{\sigma^2\Delta t^2}{m}\left(\frac{4}{c}-\frac{1}{3}\right) 
\sum_{j\neq i,\alpha}\left(  
\frac{w(r_{ij})}{r_{ij}}(r_{i\alpha}-r_{j\alpha}) 
\frac{\partial}{\partial r_{i\alpha}} \Big(  
\frac{w(r_{ij})}{r_{ij}}(r_{i\beta}-r_{j\beta})\Big)\right).
\end{equation}
For a fixed couple of indices $i$ and $j$, the differentiation takes the
form $\mathbf{X}\cdot\nabla \mathbf{X}$, with 
$\mathbf{X}=[w(r_{ij})/r_{ij}]\br_{ij}$. A further simplification comes, using
the identity\\
$\mathbf{X}\cdot\nabla \mathbf{X} =\nabla(\mathbf{X}^2)/2
-\mathbf{X}\wedge(\nabla\wedge\mathbf{X})$,
\begin{eqnarray}
\nabla_i\wedge\left(\frac{w(r_{ij})}{r_{ij}}(\br_{i}-\br_{j})
\right)
&=& 
\nabla_i\left(\frac{w(r_{ij})}{r_{ij}}\right)
\wedge(\br_{ij})
+\left(\frac{w(r_{ij})}{r_{ij}}\right)\wedge(\nabla_i\wedge\br_{ij})
\nonumber\\
&=& 0.
\end{eqnarray}
Finally, the correction reduces to 
\begin{equation}
\frac{\sigma^2\Delta t^2}{2m} \left(\frac{4}{c}-\frac{1}{3}\right) 
\sum_{j\neq i,\alpha}\frac{\partial}{\partial r_{i\alpha}}
\Big(w(r_{ij})^2\Big).
\label{eq:correctionTermDPDFinal:2}
\end{equation}
Surprisingly, the term cancels out if $c=12$, meaning that there is no
contribution at all in the case of the splitting $e^{-A/2}e^{-B}e^{-A/2}$, and
that the algorithm~(\ref{eq:splittingDPDABA:R}, \ref{eq:splittingDPDABA:P}) is
already a genuine weak order two algorithm for DPD. 

However, the correction term must be accounted for  when the splitting
$e^{-B/2}e^{-A}e^{-B/2}$ is considered, ending with the weak order two
algorithm~(\ref{eq:algoDPDBAB:R}, \ref{eq:algoDPDBAB:P},
\ref{eq:BABCorrectiveTerm}). 


\clearpage


\begin{thebibliography}{10}

\bibitem{RicCic2003}
Andrea Ricci and Giovanni Ciccotti.
\newblock Algorithms for Brownian dynamics.
\newblock {\em Molecular Physics}, 101(12):1927--1931, June 2003.

\bibitem{Ermak}
D.L. Ermak and H.~Buckholtz.
\newblock Numerical integration of the Langevin equation: Monte-carlo
  simulation.
\newblock {\em J. Comput. Phys.}, 35:169--182, 1980.

\bibitem{AllenTildesley}
M.P~Allen and D.J.~Tildesley.
\newblock {\em Computer Simulations of Liquids}.
\newblock Oxford Science Publications, Oxford, 1987

\bibitem{dunweg}
Thomas Soddeman, Burkhard D\"unweg, and Kurt Kremer.
\newblock Dissipative particle dynamics: A useful thermostat for equilibrium
  and nonequilibrium molecular dynamics simulations.
\newblock {\em Physical Review E}, 68:046702, 2003.

\bibitem{shardlow}
Tony Shardlow.
\newblock Splitting for dissipative particle dynamics.
\newblock {\em SIAM J. Sci. Comput.}, 24:1267--1282, 2003.

\bibitem{lesfinlandais}
P.~Nikunen, M.~Karttunen, and Vattulainen I.
\newblock How would you integrate the equations of motion in dissipative
  particle dynamics simulations ?
\newblock {\em Computer Physics Communication}, 153:407--423, 2003.

\bibitem{2006_Pagonabarraga_Frenkel}
I.~Pagonabarraga, Hagen M.H.J., and D.~Frenkel.
\newblock {\em Europhysics Letters}, 42:377, 1998.

\bibitem{GrootWarren1997}
Robert D.~Groot and Patrick B.~Warren.
\newblock Dissipative particle dynamics: Bridging the gap between atomistic
and mesoscopic simulation
\newblock {\em Journal of Chemical Physics}, 107(11):4423--4435, 1997.

\bibitem{remAllenTildesley}
In Ref.~\protect\cite{Ermak}, the potential force is assumed constant or slowly
varying. The extension of this method to non constant forces is presented in
Ref.~\protect\cite{AllenTildesley}, with references to M.P.~Allen.
\newblock {\em Mol.Phys}, 40:1073-1087, 1980.  
\newblock {\em Mol.Phys}, 47:599-601, 1982. and
W.F.~van Gusteren and H.J.C~Berendsen.
\newblock {\em Mol.Phys}, 45:637-647, 1982.
 
\bibitem{RemarkIto}
In It\^o calculus, the differentiation rules are different and involve second
derivatives in eq.~\protect(\ref{eq:LiouvillianDeterministic}).

\bibitem{BurBur99}
K~Burrage and P.M Burrage.
\newblock High strong order methods for non-commutative stochastic ordinary
  differential equation systems and the magnus formula.
\newblock {\em Physica~D}, 133:34, 1999.

\bibitem{FabSerEspCov2006}
G.~De~Fabritiis, M.~Serrano, P~Espanol, and P.V. Coveney.
\newblock Efficient numerical integrators for stochastic models.
\newblock {\em Physica~A}, 361(2):429, 2006.

\bibitem{SerFabEspCov2006}
M.~Serrano, G.~De~Fabritiis, P~Espanol, and P.V. Coveney.
\newblock A stochastic Trotter integration scheme for dissipative particle
dynamics. 
\newblock {\em Mathematics and Computers in Simulation}, 72:190, 2006.

\bibitem{HooKoe1992}
P.J. Hoogerbrugge and J.M.V.A. Koelman.
\newblock {\em EuroPhysics Letters}, 19:155, 1992.

\bibitem{HooKoe1993}
P.J. Hoogerbrugge and J.M.V.A. Koelman.
\newblock {\em EuroPhysics Letters}, 21:363, 1993.

\bibitem{EspWar1995}
Pep Espanol and Patrick Warren.
\newblock {\em Europhysics Letters}, 30:191, 1995.

\bibitem{Espanol1995}
Pep Espanol.
\newblock Hydrodynamics from dissipative particle dynamics.
\newblock {\em Physical Review E}, 52(2):1734, 1995.

\end{thebibliography}

\clearpage
\section*{Captions}

FIG~1. Illustration of the Fokker-Planck evolution in phase space.\\

FIG~2. Respective efficiencies of the Ermak and Ricci-Ciccotti
  algorithms. See text for details. \\

FIG~3. Pair correlation function $g(r)$ for the ideal gas and a standard
  weight function $w(r)=1-r/r_c$. From left to right, $\Delta t$ takes the
  values 0.01, 0.05 and 0.1. The algorithms presented here are ABA, BAB, BAB
  deprived of its correcting term, Shardlow S1 and DPD-VV. Simulations involve
  4000 particles, a run length of 60000 time steps, and parameters values
  $\gamma=4.5$, $T=1.$, $\sigma=3.$\\

FIG~4. Thermalization properties of ABA, BAB, Shardlow S1, DPD-VV (singular
  weight $w(r)$), ABA and DPD-VV (regular weight $w_1(r)$).  On the left
  panel: simulated temperatures $\mean{T}$  \textit{vs} time step $\Delta t$;
  on the right panel: a semi-log plot of the  difference $|\mean{T}-1|$
  \textit{vs} the time step $\Delta t$.\\ 
  
FIG~5. Pair correlation function $g(r)$ for an interacting fluid, a weight
  function $w(r)=1-r/r_c$ and a pair interaction potential $aw^2(r)/2$, with
  $a=25$. From left to right, we show the algorithm Shardlow S1, ABA and
  DPD-VV. The full line represents the result for $\Delta t=0.01$, the dashed
  line for $\Delta t=0.05$, and the symbols $\circ$ for $\Delta t=0.1$. All
  three algorithms coincide in the limit $\Delta t=0.01$. The ABA algorithm is
  more accurate for $\Delta t=0.1$ than the two others, a feature that might
  be explained by its order two nature. In the presence of repulsive
  interaction, the singularity of $w(r)$ has no consequences.\\

FIG~6. Presentation of the DPD diffusion coefficient $D(\Delta t)$ for an
  interacting fluid, for three different values of the time step $\Delta t$.
  Left panel: For each algorithm, the curves are normalized by their value at
  $\Delta t=0,01$, and only relative variations of $D(\Delta t)$ are shown, as
  in Ref.~\protect\cite{lesfinlandais}. Right panel: the bare $D(\Delta t)$
  are shown.  Depending on the algorithm chosen, the self-diffusion
  coefficient changes by a proportion of 6\%, which is also our estimate of
  the statistical error affecting our data. We conclude that the values
  obtained with $\Delta t=0.05$ are consistent with the reference values
  obtained with $\Delta t=0.01$, and that there is no significant difference
  between these three algorithms, regarding the self-diffusion properties. For
  completeness, we calculated the self-diffusion coefficient for the ideal
  gas, and found that the data obtained for the three algorithms were almost
  indistinguishable (all within $0.5\%$). The self-diffusion coefficient of
  the ideal gas is close to 0.535. The interaction potential reduces only
  lightly this value, excluding any caging effect or thermal activation.\\

FIG~7. Pair correlation function $g(r)$ for the ideal gas and a weight
  function \hbox{$w_1(r)=(r/r_c)\times(1-r/r_c)$}. From left to right, $\Delta
  t$ takes the values 0.01, 0.05 and 0.1. The algorithms presented here are
  ABA, Shardlow S1 and DPD-VV. Simulations involve 4000 particles, a run
  length of 60000 time steps, and parameters values $\gamma=4.5$, $T=1.$,
  $\sigma=3.$ 

\clearpage
\pagestyle{empty}
\section*{Figures}

\clearpage
\pagestyle{empty}

\clearpage
\pagestyle{empty}

\clearpage
\pagestyle{empty}

\clearpage
\pagestyle{empty}

\clearpage
\pagestyle{empty}

\clearpage
\pagestyle{empty}

\end{document}